\documentclass[12pt]{article}
\usepackage{latexsym,amssymb,amstext,amsmath,amsbsy}
\usepackage{hyperref}
\usepackage{bbold}
\renewcommand{\Large}{\large} 
\renewcommand{\baselinestretch}{1.2}
\allowdisplaybreaks
\def\Slash#1{\rlap{\hbox{$\mskip 3 mu /$}}#1}      
\newcommand{\ft}[2]{{\textstyle\frac{#1}{#2}}}
\renewcommand{\theequation}{\thesection.\arabic{equation}}
\renewcommand{\a}{\alpha}
\newcommand{\bs}{\boldsymbol}
\renewcommand{\b}{\beta}

\usepackage{chngcntr}
\counterwithin*{equation}{section}

\newcommand {\cA}{{\cal A}}
\newcommand {\cB}{{\cal B}}
\newcommand {\cC}{{\cal C}}
\newcommand {\cD}{{\cal D}}
\newcommand {\cE}{{\cal E}}
\newcommand {\cF}{{\cal F}}

\newcommand {\cL}{{\cal L}}

\newcommand {\cN}{{\cal N}}

\newcommand {\cR}{{\cal R}}
\newcommand {\cS}{{\cal S}}

\newcommand {\cV}{{\cal V}}

\newcommand {\cX}{{\cal X}}

\def\a{\alpha}
\def\b{\beta}

\def\t{\tau}

\newcommand{\1}{{\underline{1}}}
\newcommand{\2}{{\underline{2}}}

\newcommand{\pa}{\partial}

\renewcommand{\theequation}{\thesection.\arabic{equation}}

\newcommand{\be}{\begin{equation}}
\newcommand{\ee}{\end{equation}}
\newcommand{\bea}{\begin{eqnarray}}
\newcommand{\eea}{\end{eqnarray}}

\newcommand{\ba}{\begin{array}}
\newcommand{\ea}{\end{array}}

\def\double #1{#1{\hbox{\kern-2pt $#1$}}}

\newcommand{\bsubeq}{\begin{subequations}}
\newcommand{\esubeq}{\end{subequations}}

\newcommand{\veps}{\varepsilon}

\newcommand{\eps}{{\epsilon}}

\newcommand{\eol}{\nonumber \\}

\begin{document}

\begin{titlepage}

\begin{center}

\vskip .3in \noindent

{\Large \bf{The Resolution of an Entropy Puzzle \\ for 4D non-BPS Black Holes}}

\bigskip

	Nabamita Banerjee $^{a}$ \footnote{nabamita@iiserpune.ac.in}, Sukruti Bansal $^{a,b}$ \footnote{sukruti.bansal@pd.infn.it} and Ivano Lodato $^{a}$ \footnote{ivano@iiserpune.ac.in}\\

       \bigskip
       $^{a}$ IISER Pune, Department of Physics, Homi Bhabha Road, Pashan, Pune, India \\
       \medskip
	   $^{b}$ Dipartimento di Fisica e Astronomia `G.~Galilei', Universit\`a di Padova​ ​and INFN, Sezione di Padova, Via Marzolo 8, I-35131 Padova, Italy

       \vskip .5in
       {\bf Abstract }
       \vskip .2in
       \end{center}
We show the equality between macroscopic and microscopic black
hole entropy for a class of four dimensional non-supersymmetric black
holes in ${\cal N}=2$ supergravity theory, up to the first subleading order in
their charges. This solves a long standing entropy puzzle for this
class of black holes. The macroscopic entropy has been computed in the presence of a newly derived higher-derivative supersymmetric invariant of
\cite{{Butter:2013lta}}, connected to the five
dimensional supersymmetric Weyl squared Lagrangian. Microscopically, the crucial role in obtaining the equivalence is played by the anomalous 
gauge gravitational Chern-Simons term.


\vfill
\eject

\end{titlepage}

\tableofcontents

\section{Introduction}
\label{sec:intro}

Black Holes are thermodynamic objects and can be described in terms of
thermodynamic quantities like temperature and entropy. 
These properties are difficult
to understand at the microscopic level using statistical mechanics as
we need a fundamental theory of quantum gravity to give the 
microscopic description of black holes. String theory, the most
promising candidate of quantum gravity, has made a lot of
 progress in this regard. The observation that the thermodynamical
 properties of black holes can be studied and quantitatively 
computed with statistical methods in string theory has brought upon an overwhelming amount of research in the field.
Supergravity theory, a low energy limit of string theory, has black
holes as classical solutions. Hence, by linking the black holes 
solutions of supergravity to the states in string theory (carrying the
same charges), one can hope to relate the (macroscopic) 
entropy of a black hole to the logarithm of the degeneracies of its (stringy) microstates.\\

The initial perfect laboratory on which this link was studied was that
of BPS
black holes and BPS saturated stringy states. In fact, the computation of
statistical or microscopic black hole entropy in string theory 
makes use of the supersymmetry invariance of the background, which highly restricts the 
possible interactions between the constituent strings and branes, and
it also protects certain quantities from the effect of 
couplings, thus allowing the entropy computed at zero coupling to be
valid even in the presence of strong couplings. Therefore, 
the entropy of BPS 
black holes has been studied quite extensively using the statistical counting approach.
Although the results obtained for the entropy of BPS black holes are highly satisfactory and have led to a deeper understanding of both black hole physics and statistical string theory, it would be much more insightful to have the equivalence between the two systems worked out in the more general context of non-BPS black holes and non-BPS stringy states. Unfortunately, on the string side of this equivalence, not many advancements have been made in a general non-supersymmetric context. Nevertheless, precise results exist for certain specific cases, and in this paper we will focus on one of them. We shall study the subleading corrections to the macroscopic entropy of extremal non-supersymmetric black holes in ${\cal N}=2$ supergravity in four dimensions and show its relation with the microscopic result. \\

The thermodynamical or macroscopic entropy of black holes can be worked out irrespective of any special symmetry. It is well known that the classical entropy of a black hole, given by Bekenstein-Hawking (BH) area law, corresponds to the area of the black hole's event horizon. Subleading corrections to BH formula, in the semiclassical ``thermodynamical limit'' of large charges, can be obtained by a method engineered by Wald \cite{Wald:1993nt} long ago. In this paper we use another, different but equivalent, method, known as the ``entropy function formalism", given by Sen \cite{Sen:2005wa}, which is well-suited to compute the macroscopic entropy of extremal black holes, and sub-leading corrections thereof. \\

The computation of semiclassical microscopic entropy of black holes,
which is the logarithm of the degeneracy of states in 
string theory, hinges strongly on supersymmetry of the theory. This entropy is 
given, in the semiclassical limit, by the Cardy formula 
\begin{equation}
\label{eq:cardy}
S_{\rm BH}=2\,\pi \Big(\sqrt{\frac{c_L\,h_L}{6}}+\sqrt{\frac{c_R\,h_R}{6}}\Big),
\end{equation}
where $c_L$ and $c_R$ are the left and right-handed central charges respectively in conformal field theory, and $h_L$ and $h_R$  are the left and right-handed zero modes of Virasoro generators respectively. Note that, since the goal is to compare the statistical result with the macroscopic one, the stringy states
must correspond to a black hole configuration.\\
 There are specific examples of string configurations which allow an exact counting of states even in the absence of supersymmetry \cite{Harvey:1998bx,Kraus:2005vz,Kraus:2005zm}. Their approach is based on a more basic/primitive 
form of the AdS/CFT correspondence \cite{Brown:1986nw} as we will explain later on.\\

The final result for the microscopic entropy of a large class of five dimensional non-supersymmetric black 
holes in ${\cal N}=2$ supergravity with subleading corrections as obtained in
\cite{Harvey:1998bx,Kraus:2005vz,Kraus:2005zm} is given by
\begin{equation}\label{KLRESULT}
S^{\rm micro}_{\rm non-BPS}=2\,\pi\sqrt{q_0(d_{ABC}\,p^A\,p^B\,p^C+128\,d_A\,p^A)}
\end{equation}
where $q_0$ is the electric charge of the black hole, $p^A$'s are the
magnetic charges, and $d_{ABC}$ and $d_{A}$ are constant 
factors appearing in the prepotential of the Lagrangian of the black
hole solution\footnote{Their physical interpretation will be given in
  section \ref{sec3.1}.}. The computation of the macroscopic 
counterpart of the statistically calculated entropy \eqref{eq:cardy}
was attempted in \cite{Sahoo:2006rp}  and the following 
result was obtained: 
\begin{equation}
S^{\rm macro}_{\rm non-BPS}=2\,\pi\,\sqrt{q_0(d_{ABC}\,p^A\,p^B\,p^C)}\Big(1+\frac{40\,d_A\,p^A}{d_{ABC}\,p^A\,p^B\,p^C}\Big)\;.
\end{equation}
Evidently, this does not correspond to a first order expansion of \eqref{KLRESULT} in the limit of large charges. The most likely reason behind this mismatch, as proposed in \cite{Sahoo:2006rp}, is connected to the form of the higher derivative interaction terms. To be precise, the Weyl squared invariant of superconformal gravity, that was considered in \cite{Sahoo:2006rp}, might not be enough to describe all the subleading contributions for a non-BPS solution. This directed us towards the possibility of certain (supersymmetric) higher derivative invariants missing in the four dimensional Lagrangian, that could be connected to five  dimensional theories considered in \cite{Kraus:2005vz,Kraus:2005zm}, and hence  readily affect the macroscopic results.\\

Recently, a new class of higher derivative invariants was built in the
context of four dimensional superconformal 
gravity \cite{Butter:2013lta}. The most important property of this new
class of invariants is its connection to the five dimensional
supersymmetrization of the Weyl squared Lagrangian, which contains the
gauge gravitational Chern-Simons term considered in
\cite{Kraus:2005vz,Kraus:2005zm}. This higher derivative invariant has
also been shown to not contribute to the entropy of BPS black holes
\cite{Butter:2014iwa}.  This was 
expected because the micro- and
macroscopic entropy results for BPS black holes already
matched\cite{Maldacena:1997de,LopesCardoso:1998tkj,LopesCardoso:1999cv} without
considering these corrections. In this paper, we consider this new higher derivative invariant to compute its contribution to the entropy of the non-BPS black hole solutions. The final entropy we obtain is given as,
\begin{equation}
S^{\rm macro}_{\rm non-BPS}=2\,\pi\,\sqrt{q_0(d_{ABC}\,p^A\,p^B\,p^C)}\Big(1+\frac{64\,d_A\,p^A}{d_{ABC}\,p^A\,p^B\,p^C}\Big)\;,
\end{equation}
and it exactly matches the first order expansion of statistical entropy in the limit of large charges \eqref{KLRESULT}, as predicted in \cite{Kraus:2005vz,Kraus:2005zm}! Hence, 
we have successfully moved one step further in describing black hole
thermodynamics in the context of string theory via the more general
case of non-BPS black holes in supergravity.\\

In the following we will closely follow \cite{Sahoo:2006rp},
focusing on a general class of
theories, the STU models. The paper is organized as follows: in section \ref{sec:4Dtheory} we present the 4D theory in its prepotential formulation and add a new class of supersymmetric higher derivative invariants, recently
discovered and analyzed in \cite{Butter:2013lta,Butter:2014iwa}. The
new higher derivative invariant was the missing ingredient in the
previous calculation worked out in \cite{Sahoo:2006rp} and it is
the key to the resolution of the puzzle. In section
\ref{sec:5Dresults} we give an outline of the microscopic entropy
computation for a specific class of black holes, and make contact with
the lower dimensional theory and solutions in four dimensions.
Finally in section \ref{sec:end} we conclude with the results and open
questions. Our conventions and the details of the 4D $\cal{N}=$2 superconformal
theory are presented in the two
appendices at the end of this paper.\\

\section{\texorpdfstring {${\cal N}=2$}{Lg} Superconformal Gravity in 4D Setting}
\label{sec:4Dtheory}
\setcounter{equation}{0}

We want to study the subleading correction to the entropy of non-BPS black
holes in four dimensional $\cN=2$ supergravity. 
The theory we use describes the dynamics of four vector multiplets
$X^I$  ($I,J=0,\dots,3$) coupled to conformal supergravity and a
chiral background multiplet ${\bs A}$, which will soon be identified
with a particular linear combination of the Weyl squared 
multiplet \cite{Bergshoeff:1980is} and the TLog
multiplet \cite{Butter:2013lta}.
This, as shown in 
\cite{Banerjee:2011ts} and proven in \cite{Butter:2014iwa}, corresponds exactly
to the dimensional reduction of the 5D higher 
derivative invariant containing the Weyl squared
term and the gravitational Chern-Simons term 
$A \wedge R \wedge R$ \cite{Hanaki:2006pj}. \\
The Lagrangian can be written in terms of a homogeneous function of
degree 2,
$F$ ($A,B=1,2,3$):
\begin{equation}
\label{eq:prepotential}
F(X^I,{\bs\cA})=- \,d_{ABC}\frac{\,X^A\,X^B\,X^C}{X^0}-d_A\,\frac{\,X^A}{X^0}\,{\bs \cA}\;,
\end{equation} 
where $X^I$ and ${\bs \cA}$ are the lowest components of the vector and
chiral multiplets of $\cN=2$ theory respectively:
\begin{equation}
{{\cal X}^I}=(X^I,\Omega^I_i, W^I_\mu,Y^I_{ij})\;,
\qquad {\bs A}=({\bs \cA},{\bs \psi_i},{\bs \cB_{ij}},{\bs \cF^-_{ab}},{\bs\Lambda_i},{\bs \cC})\;.
\end{equation}
We refer to appendix \ref{app.B} for a more detailed discussion on the
supersymmetry transformation rules of these multiplets. 
For now it suffices to say that they are representations of
the full superconformal algebra ${\rm SU}(2,2|2)$, which, 
in this particular example, corresponds to the algebra of the symmetries
of the Lagrangian\footnote{R-symmetry is not 
necessarily a symmetry of the action.}. For simplicity, we will focus
only on the bosonic sector of the action, which 
reads \footnote{Our conventions
  differ from \cite{Sahoo:2006rp} by a minus sign in the Riemann
  tensor and we write down the explicit dependence on D in the action.} (refer to the appendices \ref{app.A} and \ref{app.B} for the definitions of the fields and derivatives used in the following):
\begin{align}
\label{eq:Lagrangian_prepot}
8 \, \pi \, \cL =& \,\, {\rm i} (X^I \bar F_I - \bar X^I F_I)(\tfrac16\, R-D)
+ \Bigg[ {\rm i}\, \cD_a F_I \cD^a \bar X^I
\nonumber \\
&
+\frac{\rm i}{4} F_{IJ} (F^{-I}_{ab} -\tfrac14 \bar X^I 
T^-_{ab})
(F^{-Jab} -\frac{1}{4} \bar X^J T^{-ab}) \nonumber \\
& +\frac{\rm i}{8} \bar F_I  (F^{-I}_{ab} -\frac{1}{4} \bar X^I 
T^-_{ab}) T^{-\,ab} - \frac{\rm i}{8} F_{IJ} Y^I_{ij}
Y^{Jij} + \frac{\rm i}{32} \bar F \, (T^-_{ab})^2
\nonumber \\
& + \frac{\rm i}{2} F_{\bs \cA} {\bs \cC} - \frac{\rm i}{8} F_{\bs \cA
\bs \cA} ({\bs \cB}_{ij} {\bs \cB}^{ij} - 2 {\bs \cF}^-_{ab}
{\bs \cF}^{-ab}) \nonumber \\
& +\frac{\rm i}{2} {\bs \cF}^{-\,ab} F_{{\bs \cA} I} 
(F^{-I}_{ab} -\frac{1}{4} \bar X^I 
T^-_{ab}) -\frac{\rm i}{4} {\bs \cB}_{ij} F_{{\bs \cA} I} Y^{Iij}
 + h.c. \Bigg]\nonumber \\
& + {\rm i} (X^I \bar F_I - \bar X^I F_I) 
\bigg( \cD^a V_a - \frac{1}{2} V^a V_a - \frac{1}{ 4}
|M_{ij}|^2    \nonumber \\
&  + (\partial^\mu \Phi^i_\alpha + \frac{1}{2}\,
\cV^{i\mu}_{~~j}
\Phi^j_\alpha) (\partial_\mu \Phi^\alpha_i + \frac{1}{2}\,
\cV^{~k}_{i\mu}
\Phi^\alpha_k)\bigg), \,
\end{align}
where the following definitions have been used:
\begin{align}
 F^I_{\mu
  \nu}=2\,\partial_{[\mu}{}W^I{}_{\nu]} \;,& \qquad
T^-_{ab}=T_{ab}{}^{ij}\,\varepsilon_{ij}\;,
\qquad T^+_{ab}=T_{ab\,ij}\,\varepsilon^{ij}\;.
\nonumber\\
F_I &= \partial_{X^I} F \;, \qquad F_{\bs \cA}=\partial_{\bs \cA} F\;, \quad\dots\;.
\end{align}
In Lagrangian \eqref{eq:Lagrangian_prepot}, $V_a$ is a vector field, $M_{ij}$ is an SU(2) triplet scalar field and $\Phi^\alpha_i$ is an SU(2) matrix valued scalar field. The indices $i, j \, (=1,2)$ and $\alpha \, (=1,2)$ used here, label the fundamental representation of gauged and global SU(2) respectively.\\
We couple the theory to a non-linear multiplet whose
gauge field $V_\mu$ is subject to the constraint:
\begin{equation}
\label{eq:non-lin_constr}
D^a\,V_a-\tfrac12\,V^a\,V_a-\tfrac14\,|M_{ij}|^2+(\partial^\mu \Phi^i_\alpha+\tfrac12\,\cV^{i\,\mu}_{\;j}\,\Phi^j_\alpha)(\partial_\mu \Phi_i^\alpha+\tfrac12\,\cV_{i\,\mu}^{\;k}\,\Phi_k^\alpha)-3D=0\;,
\end{equation}
which reduces the number of its independent degrees of freedom from four to
three, balancing the number of independent bosonic and fermionic fields. More importantly it
allows to fix the value of the auxiliary
field $D$ in terms of the Ricci scalar $R$. 
Note that $V_\mu$ transforms under K-boosts, so its 
covariant derivative can be written as $D_a V^a=\cD_a V^a -2\,f_a{}^a=\cD_a V^a-\tfrac13\,R+2D$.\\

We are now ready to identify the chiral background with the two higher
derivative invariants that come from the dimensional 
reduction of the $R^2$ term in 5D $\cN=2$ supergravity, the Weyl squared invariant and the TLog multiplet, i.e \cite{Butter:2014iwa}
\begin{equation}
\label{eq:full_prepot}
{\bs \cA}=(T^-_{ab})^2-\tfrac{32}3\,A|_{\mathbb{T}(\ln\,\bar X^0)}\;.
\end{equation}
Although the Weyl squared invariant depends 
only on conformal supergravity curvatures and auxiliary fields, the
TLog multiplet, as it comes from the dimensional reduction 
from 5D, depends explicitly on the compensating vector multiplet
$\cX^0$. Of course, it would seem logical (but nevertheless 
incorrect) to consider the dependence of  ${\bs A}$ on $\cX^0$ ,
while taking the derivatives of the prepotential wrt the scalar 
fields $X^I$.  The
multiplets at this stage must be considered elementary (independent). This 
means, for instance, that the derivative of the prepotential wrt to
$X_0$ is simply given by $F_0= d_A\,\frac{X^A}{X_0^2}\, {\bs \cA}$. 
Once the supersymmetric Lagrangian has been obtained, the components
of ${\bs A}$ can be traded with their composite 
expressions.
Neglecting all the fermions, these components read (the indices $a,b=0,\dots,3$ are flat or
tangent space indices)
\begin{align}
  \label{eq:high_der_components}
{\bs \cA}=& \,(T^-_{ab})^2-\tfrac{32}3\,\Big(-2\,\frac{\Box_\mathrm{c} X^0}{\bar X^0}-\tfrac14\,\frac{\cF^{-\,0}_{ab}\,T^{-\,ab}}{\bar X^0}+\frac{1}{4\,(\bar X^0)^2}\Big(Y^{0\,ij}\,Y^0_{ij}-2\,\cF^{+\,0}_{ab}\,\cF^{+\,0\,ab}\Big)\Big)
\,, \nonumber\\[.3ex]
 {\bs  \cB_{ij}}=& -16 \,\varepsilon_{k(i}R({\cal
    V})^k{}_{j)ab} \, T^{-\,ab}\,
	+\tfrac{64}3\,\big(\Box_\mathrm{c} + 3\,D\big) \frac{Y^{ij}}{\bar X^0}
	+\tfrac{64}3\, \frac{\cF^{+\,0}_{ab}}{\bar X^0}\, R(\mathcal{V})^{ab\,k}{}_{i}\, \varepsilon_{jk} \,,
	\nonumber \\[.6ex]
   {\bs \cF_{ab}^-} =&-16 \,\cR(M)_{cd}{}^{\!ab} \,
  T^{-cd}\,-\tfrac{32}3 \Box_\mathrm{c} \log\bar X^0 \,T^-_{ab}
	+\tfrac{32}3 R(\mathcal{V})^-{}_{\!\!ab}{}^i{}_k \,\frac{Y^{0\,jk}}{\bar X^0} \,\varepsilon_{ij}
	- \tfrac23 T^-_{ab}\,T^+_{cd}\frac{\cF^{+\,0\,cd}}{\bar X^0}
	\nonumber\\ 
        &  \,+\tfrac{32}3 \big(\delta_a{}^{[c} \delta_b{}^{d]}
    - \tfrac12\varepsilon_{ab}{}^{cd}\big)
		\big[4\, D_c D^e \frac{\cF^{+\,0}_{ed}}{\bar X^0}
		+ (D^e \log\bar X^0\,D_c T^-_{de}
		+ D_c \log\bar X^0 \,D^e T^-_{ed})
		- w D_c D^e T^-_{ed}
	\big] \,, 	
	\nonumber\\[.6ex]       
{\bs \cC} =& \,64\, \cR(M)^{-cd}{}_{\!ab}\,
 \cR(M)^-_{cd}{}^{\!ab}  +32 R({\cal V})^{-ab\,k}{}_l^{~} \,
  R({\cal V})^-_{ab}{}^{\!l}{}_k  -16 T^{-\,ab} \, D_a \,D^c T^+_{cb}\,
  \nonumber\\
  &\,-\tfrac{128}3 (\Box_{\rm c} + 3 D)
        \Box_{\rm c} \log\bar X^0
	-64 (D_a D) \, D^a \log\bar X^0
	+\tfrac{512}3 \,D^a \Big(R(D)_{ab}^+\, D^b \log\bar X^0\Big)
	\eol & \,
	+\tfrac{16}{3}\, D^a (T^+_{ab}\, T^{-\,cb} D_c \log\bar X^0)
	+ \tfrac{8}{3} D^a(T^+_{ab}\, T^{-\,cb}) D_c \log\bar X^0
	\eol & \,
	- \tfrac{16}{3} D_a D^a (T^+_{bc}\, \frac{\cF^{+\,0\,bc}}{\bar X^0})
	- \tfrac{64}3 \,D_a \Big(D^b T^+_{bc}\, \frac{\cF^{+\,0\,ac}}{\bar X^0}+ D^b
       \frac{\cF^{+\,0}_{bc}}{\bar X^0}\, T^{+\,ac}\Big) - \tfrac{2}{3} (T^+_{ab})^2 A\vert_{\mathbb{T}(\log\bar X^0)}
	\eol & \,
        - \tfrac{32}3\,w\,\Big\{
	- R(\cV)_{ab}^+{}^i{}_j R(\cV)^{ab+}{}^j{}_i
	- 8  R(D)^+_{ab} R(D)^{ab+}
	\eol &\quad \qquad
	- \tfrac12 D^a T^+_{ab}\, D_c T^{-\,cb}
	- \tfrac12 D^a (T^+_{ab}\, D_c T^{-\,cb})\Big\} \,,
\end{align}
where $\cF^{-\,0}_{ab}=F^{-\,0}_{ab}-\tfrac14\,\bar X^0\,T^-_{ab}$ and
we made use of the constraints relating the components of a chiral
multiplet to those of a vector multiplet (see \eqref{constrvect2}).
Since we have fixed the TLog multiplet to be a composite function of the
vector multiplet $\cX^0$, the parameter $w$ above is fixed to one \cite{Butter:2013lta}.

\subsection{The Background and the Auxiliary Field}
\label{subsec:background}
We want to study the entropy of extremal, non-BPS black holes. The near horizon geometry of an extremal black hole in four dimensions is described by ${\rm AdS}_2\times {\rm S}^2$  metric, given by
\begin{equation}
\label{eq:metric}
ds^2=v_1\bigg(-r^2\,{\rm d}t^2+\frac{{\rm d}r^2}{r^2}\bigg)\,+v_2\,(\,{\rm d}\theta^2+\sin^2\theta\,{\rm d}\phi^2\,)\,.
\end{equation}
Here $v_1$ and $v_2$ are the scaling parameters of the $\rm AdS_2$ and
$\rm S^2$ spaces respectively. The invariance under 
the symmetry group ${\rm SO}(2,1)\times {\rm
  SO}(3)$ of the ${\rm AdS}_2\times {\rm S}^2$  metric must be satisfied by all
field configurations. As an example, in flat space indices, the field strengths 
will have only two non-zero components, one along $(0,1)$ direction and
the other along $(2,3)$ direction as,
\begin{equation}
\label{eq:field_strenghts}
F^I_{01}=-\frac{e^I}{v_1}\;,\quad F^I_{23}=\frac{p^I}{v_2} \,,
\end{equation}
where $e^I$'s are the generalized electric fields, $p^I$'s are the magnetic charges of the black hole. The electric charges are defined via the dual field strength:
\begin{equation}
G^-_{I\,ab}=(-2\,{\rm i})\,\frac{\partial (8\pi\,\cL)}{\partial F^{-\,I\,ab}}
\end{equation}
and are given by:
\begin{equation}
\frac{q_I}{v_2}=G_{I\,23}\;.
\end{equation}
Analogously the auxiliary $T^\pm_{ab}$ tensor will have only two
non-zero components, one on the subspace $(0,1)$ and the 
other in the subspace $(2,3)$,
\begin{equation}
\label{eq:T_tensor}
T^-_{01}={\rm i}\, T^-_{23}= - z\;,
\end{equation}
with $z$ complex constant. Another important quantity, which is completely fixed by the symmetries of
the background, is the modified Lorentz curvature 
$\cR(M)_{ab}{}^{cd}$. Here we present the non-zero components (from
here onwards we take $R(A)=0$ for reasons that will be explained shortly),
\begin{align}
  \label{eq:R(M)-values}
  \mathcal{R}(M)_{\underline{a} \underline{b}}{}^{\underline{c}
    \underline{d}} =&\,(D+\ft13 R)\, \delta_{\underline{a}\underline{b}}
  {}^{\underline{c}\underline{d}}  \,,\nonumber\\ 
  \mathcal{R}(M)_{\hat a \hat b}{}^{\hat c \hat d} =&\,(D+\ft13 R)\,
  \delta_{\hat a\hat  b}{}^{\hat c\hat d}\,,\nonumber\\ 
  \mathcal{R}(M)_{\underline{a} \hat b}{}^{\underline{c} \hat d}
  =&\,\ft12(D-\ft16 R)\, \delta_{\underline{a}}{}^{\underline{c}} \,
  \delta_{\hat b}{}^{\hat d} \,,
\end{align}
where the underlined and hatted indices span the $(t,r)$ and $(\theta ,\phi)$
subspaces respectively. Now, by using the definition of 
$\cR(M)_{ab}{}^{cd}$ in terms of the Weyl tensor $C_{ab}{}^{cd}$,
\begin{equation}
\cR(M)_{ab}{}^{cd}=C_{ab}{}^{cd}+D\,\delta_{[a}^c\,\delta_{b]}^d \, ,
\end{equation}
we obtain
\begin{equation}
\label{eq:full_trace_R(M)pm}
\cR(M)^{\pm\;\;ab}_{ab}=\tfrac12 \cR(M)_{ab}{}^{ab}=3D \,.
\end{equation}
There are many other (non-)dynamical fields attracted on the horizon,
and in principle one would need to solve their equations of 
motion to find their constant values explicitly. For simplicity, we will consider 
 the consistent truncation of the full field configuration, satisfying
 the near-horizon symmetries and describing (non-)BPS black holes 
in $\cN=2$ supergravity, presented in \cite{Sahoo:2006rp}. 
In the following we shall sketch the truncation procedure, proving its consistency. Although for the purpose of this paper we need consistent solutions of the two derivative
theory only, the truncation remains consistent even after the addition of higher
derivative invariants. \\ 

Let us start from the auxiliary fields of the vector multiplet, in
particular $Y^I_{ij}$. The (full) equations of motion for these four triplets
are quadratic in the $Y^I$'s, so a simple consistent solution of the dynamical equations is $Y^I_{ij}=0$. Next, we analyze the
 auxiliary gauge fields, the R-symmetry connections of the Weyl
 multiplet  $A_\mu$ and 
$\cV_{\mu \;\; j}^{\;i}$ (see appendix \ref{app.B}). We first note that, if $A_\mu$ is
constant, its curvature $R(A)=0$. This means that locally $A_\mu$
can be taken to be vanishing. Furthermore, the vanishing of the ${\rm
  U}(1)_{\rm R}$ curvature implies $R(D)=0$ (eq. \eqref{eq:RM-constraints}) which implies that also the dilatational gauge $b_\mu$ can also be taken
to be locally vanishing. On the other hand when the ${\rm SU}(2)_{\rm R}$ connection $\cV_{\mu \;\; j}^{\;i}$ is constant, the curvature $R(\cV)_{\mu\nu\;\;j}^{\;\;i}\sim\cV_\mu\,\cV_\nu$. It is easy to check by inspection that the (higher derivative) equations of motion for $\cV_{\mu \;\; j}^{\;i}$ are quadratic or quartic in it, so they also admit the solution $\cV_{\mu \;\; j}^{\;i}=0$. The fields of the compensating non-linear multiplet can be trivially fixed through 
their equations of motion. The ${\rm SU}(2)$ triplet $M_{ij}$ can be fixed to zero, and the field $\Phi^i_\alpha$, obeying the constraints $\Phi^\dagger \Phi=\mathbb{1}$ and $\det \Phi=1$, can be fixed to the constant $\delta^i_\alpha$. The vector $V_\mu$ is constrained by eq. \eqref{eq:non-lin_constr}, and admits the simple solution $V_\mu=0$. This in turn leads to a
constraint on the auxiliary field $D$, i.e. $D+\tfrac13\,R=0$.\\
 The full near-horizon configuration at the two derivative level is then:
 \begin{align}
 \label{eq:near_horizon_conf_1}
 T^-_{01}&=-z\;,\quad & A_\mu&=\cV_{\mu \;j}^{\;\; i}=b_\mu=V_\mu=M_{ij}=0\;,
 \nonumber\\
 \Phi^i_\a&=\delta^i_\alpha\;,\quad & D&=-\tfrac13\,R \,.
 \end{align}
We want to point out that on this background all the fields are covariantly constant because
the symmetries of the geometry allow for spin connections orthogonal to the field
configurations. For e.g., one can check explicitly that $\cD_\mu T^\pm_{ab}=0$.
Note that the addition of the TLog multiplet brings about terms linear
in the ${\rm SU}(2)_R$ connection through the covariant 
derivative of the auxiliary field $Y^0_{ij}$ of the vector multiplet
$\cX^0$. As we pointed out before, however, the auxiliary 
fields $Y^I_{ij}=0$, and so the linear dependence on $\cV_{\mu \; j}^{\;\; i}$ is removed.
An equivalent approach would have been to \emph{assume} all the fields
to be covariantly constant and find, as a solution, 
$A_\mu=b_\mu=\cV_{\mu \; j}^{\;\; i}=0$.\\
Finally we note that the auxiliary real scalar $D$ does
 not appear in the Lagrangian at the two derivative level, so in 
principle, it cannot be fixed. However, since it enters the non-linear
multiplet constraint, its most general consistent configuration is
indeed $D=-\tfrac13\,R$ even after 
 the higher derivative corrections 
are considered. But, as will become clear shortly, we will not
  need the near-horizon field configurations beyond the 
leading two derivative order, to compute the first order subleading
corrections to the entropy of the (non)-BPS black hole 
solutions.

\subsection{The Entropy Function}
In this section we set up the computation of the moduli and the
entropy of a class of black holes with near-horizon geometry 
\eqref{eq:metric}. We will use Sen's entropy function formalism for this purpose.
The entropy function is given as
\begin{equation}
\cE(v_1,v_2,z,X^I,e^I,q_I,p^I)=2\,\pi\,\Big(-\tfrac12\,e^I q_I-\int{\rm d}\theta{\rm d}\phi\,\sqrt{-\det g}\,\cL\Big)\;.
\end{equation}
This function need to be evaluated for the background described by \eqref{eq:near_horizon_conf_1}. Hence 
$\cE$ will be a function of the charges and various near horizon
parameters. The entropy function formalism requires the entropy function to be extremized with respect to the near horizon parameters, i.e. obey the following extremization equations: 
\begin{equation}
\label{eq:extremization_eq}
\frac{\partial \cE}{\partial v_{1,2}}=\frac{\partial \cE}{\partial X^I}=\frac{\partial \cE}{\partial e^I}=\frac{\partial \cE}{\partial z}=0 \,.
\end{equation}
The values of the moduli will either remain unfixed , as is the case
for the auxiliary field $D$ at the two derivative level, or will get 
fixed in terms of the electric and magnetic charges $q_I$ and
$p^I$. 
The extremum value of $\cE$ by definition gives us the entropy of the black hole, i.e. 
\begin{equation}
S_{\rm BH}=\cE\vert_{v_{1,2},z,X^I,e^I}\,.
\end{equation}
Before we proceed to adapt the entropy function formalism for our purpose, we
want to show how it simplifies the computations while treating the higher derivative
Lagrangians perturbatively. Consider the higher derivative perturbative
effective actions to be specified by a 
small parameter $\delta$.  In their presence, various near horizon parameters will get corrections proportional to
$\delta$. To be precise,
let $\Phi$ denotes all possible near horizon parameters
$(v_1,v_2,z,X^I,e^I,....)$. In presence of perturbative higher
derivative interactions, the entropy function can be 
schematically written as $\cE=\cE_0+\delta \cE_1$ and the perturbative
solutions of the near horizon parameters as $\Phi=\Phi_0+\delta\cdot
\Phi_1$. Here $\cE_0$ is the entropy function for two derivative
theory and $\Phi_0$ is the solution of the corresponding extremization
equations. Finally, the value of the extremized entropy function to
first order in $\delta$ looks like,
\begin{equation}
\cE(\Phi)=\cE_0(\Phi_0+ \delta \Phi_1)+\delta \cdot\,\cE_1(\Phi_0)= \cE_0(\Phi_0)+\delta \cdot\,\cE_1(\Phi_0) \,.
\end{equation}
The dependence on $\Phi_1$ will vanish identically by the leading
order extremization condition. This means, to compute the
first subleading correction to the entropy of black holes, 
using the entropy function formalism, we will need only the two
derivative solutions for the near horizon parameters. 
Hence, as hinted above, for computing non-BPS black holes entropy
accurate up to the first order in $\delta$, i.e. due to the first order effective corrections of the theory, we will 
only need the consistent solutions for near horizon
parameters presented in \eqref{eq:near_horizon_conf_1}. \\

Now, plugging the near horizon configuration of section \ref{subsec:background} into the entropy function for our case, we obtain:
\begin{align}
\label{eq:g_prepot}
\cE&\equiv -\pi\,q_I\,e^I -\pi g(v_1,v_2,z,X^I,e^I,p^I)
\nonumber\\
&=
-\pi\,q_I\,e^I -\pi\Big\{{\rm i}\,(v_2-v_1)\big(X^I\,\bar F_I- \bar X^I\,F_I\big)
\nonumber\\
&\,-\Big[\frac{\rm i}{4}\,v_2\,v_1^{-1}\,F_{IJ}\big(e^I-{\rm i}\,\frac{v_1}{v_2}\,p^I-\tfrac12\,\bar X^I\,z\,v_1\big)\big(e^J-{\rm i}\,\frac{v_1}{v_2}\,p^J-\tfrac12\,\bar X^J\,z\,v_1\big)+\mathrm{h.c.}\Big]
\nonumber\\
&\,-\Big[\frac{\rm i}{4}\,\bar F_I\,z\,v_2\big(e^I-{\rm i}\,\frac{v_1}{v_2}\,p^I-\tfrac12\,\bar X^I\,z\,v_1\big)+\mathrm{h.c.}\Big]+
\Big[\frac{\rm i}{8}\,F\,\bar z^2\,\,v_1\,v_2+\mathrm{h.c.}\Big]
\nonumber\\
&\,+\Big[32\,{\rm i}\, F_{\bs \cA}\Big(\,\frac{8}{3}\frac{v_1}{v_2}+\frac{8}{3}\frac{v_2}{v_1}-4+\frac{7}{192}\,v_1\,v_2\,|z|^4-\frac{1}{3}\,|z|^2\,(v_1+v_2)\Big)+{\rm h.c.}\Big]
\nonumber\\
&\,+\Big[\frac{4\,\rm i}{3}F_{\bs \cA}\,v_1\,v_2\,{\bar z}^2\,A|_{\mathbb{T}(\ln\,\bar X^0)}+{\rm h.c.}\Big]
-\Big[\frac{16\,z}{3\,v_1}(v_1-v_2)\,F_{{\bs \cA}\,I}\Big(e^I-{\rm i}\,\frac{v_1}{v_2}\,p^I-\tfrac12\,v_1\,z\,\bar X^I\Big)+{\rm h.c.}\Big]
\nonumber\\
&\,-\Big[\,\frac{32}{3}\frac{1}{\bar X^0}\frac{v_2}{v_1}\,F_{{\bs \cA}\,I}\Big(e^I-{\rm i}\,\frac{v_1}{v_2}\,p^I-\tfrac12\,v_1\,z\,\bar X^I\Big)\Big(e^0+{\rm i}\,\frac{v_1}{v_2}\,p^0-\tfrac12\,v_1\,\bar z\,X^0\Big)\Big(-\frac1{v_1}-\frac1{v_2}+\frac14\,|z|^2\Big)+{\rm h.c.}\Big]\Big\}.
\end{align}

In the next section we will focus on two particular classes of solutions: fully BPS and non-BPS black holes, and we will compute their entropy by extremizing the entropy function $\cE$. Before doing so, we will review the microscopic/statistical entropy results, based on anomalies.

\section{Entropy of 5D/4D (non-)BPS Black Holes}
\label{sec:5Dresults}
As mentioned earlier, although the computation of non-supersymmetric indices is still not well understood, it is possbile to obtain quantitative results for the statistical entropy of non-BPS black holes in very specific cases. In the following we will give a short summary of the results obtained by \cite{Harvey:1998bx,Kraus:2005vz,Kraus:2005zm} for five-dimensional supergravity black holes, stressing the crucial ingredients. It is worth stressing that identical results were already presented in \cite{Maldacena:1997de}, but there the interest was focused on BPS solutions. We will also discuss the connection to four-dimensional black hole solutions of $\cN=2$ supergravity, relevant to this work. \\

\subsection{Statistical Entropy of (non-)BPS Black Holes in 5D: A Sketch}\label{sec3.1}
Consider a near-horizon black brane configuration 
of the form ${\rm AdS}_3\times S^2\times X$, where $X$ is a Calabi-Yau
three-fold, that arises from M-theory \footnote{A well-known example
  of such a compactification is the case when $X=K3\times 
T^2$. An $M5$-brane on such a manifold is dual to heterotic string
theory, which has both a susy and a non-susy sector. Our 
results, as we shall see, are sensitive to all the excitations
independent of their supersymmetries.}.  After compactifying 
on $X$, we get the near-horizon solution of a five dimensional black
hole, which, depending on the near-horizon values of the 
matter and gravity fields, may or may not preserve
supersymmetries. The theory can
be further reduced over a two-sphere to obtain a solution of three dimensional
supergravity, which includes the dimensionally reduced 
Chern-Simons form $A\wedge {\mathrm d}A$. A crucial point is that
the left and right central charges of the CFT that lives at the  
boundary of ${\rm AdS}_3$ can now be computed from the anomalous
terms that are connected to the Chern-Simons terms of the reduced 5D
supergravity Lagrangian. The procedure requires supersymmetry of
the theory, but not necessarily of the background. Hence 
we will obtain a result for non-BPS configurations as well.
In this section we want to sketch the main results of the anomaly
analysis, referring to
\cite{Harvey:1998bx,Kraus:2005vz,Kraus:2005zm,Maldacena:1997de,Freed:1998tg,Sahoo:2006vz}
and the references therein for more detailed discussions. \\

The first goal of this section is to obtain the Cardy formula for
the entropy of black holes whose near-horizon 
configuration contains an 
${\rm AdS}_3$ factor \cite{Brown:1986nw}. Note that there
are certain advantages in considering an ${\rm AdS}_3$ 
near-horizon configuration instead of ${\rm AdS}_2$, since the spacetime
symmetries of the former close into the infinite dimensional Virasoro algebra.\\

One way to compute the black hole entropy is to consider the Euclidean
continuation of the solution. One then
obtains the thermodynamics potential from which it is straightforward
to compute the entropy. In Euclidean signature, the 
BTZ black hole is a solid torus having 
one contractible and one non-contractible cycle. We denote the coordinates
along these cycles respectively as $t_{C}$ and $t_{NC}$. There is an entire family of
solutions that can be derived from the Euclidean one by 
choosing to identify the time coordinate with a certain combination of
$t_{NC}$ and $t_C$. For our purpose it is enough to 
consider just the two cases, $t_{NC}\rightarrow -{\rm i}\,t$ and
$t_{C}\rightarrow -{\rm i}\, t$. In the former case, one obtains the 
geometry of global ${\rm AdS}_3$ with compact imaginary time
corresponding to thermal AdS.  The latter choice leads instead
to the BTZ black hole. Note that the coordinates
$t_{NC}$ and $t_C$ are related by a simple modular transformation 
of the boundary torus, $\t\rightarrow -\tfrac1{\t}$. Hence if we obtain
the partition function for one such solution, it is immediate 
to obtain the partition function for the other one.\\

For a generic asymptotically ${\rm AdS}_3$ solution, the conserved
quantities are the energy $H$ and the angular momentum $J$, 
which corresponds to the momentum in the CFT at the boundary. The partition function then reads:
\begin{equation}
Z(\b,\mu)=e^{-S_E}={\rm Tr}\,e^{-\b\,H-\mu\,J}={\rm Tr}\,e^{2\,\pi\,{\rm i}\,\t\,h_L}\,e^{-2\,\pi\,{\rm i}\,\bar\t\,h_R} \,,
\end{equation}
where the following definitions have been used for the modular parameter of the torus, $\t$, and the zero modes of the Virasoro algebra:
\begin{align}
\label{eq:tau_h}
\t &={\rm i}\,\frac{\b-\mu}{2\,\pi}\;, & \bar\t &=-{\rm i}\,\frac{\b+\mu}{2\,\pi}\;,
\nonumber\\
h_L &= L_0-\frac{c_L}{24}=\frac{H-J}{2}\;, & h_R &=\tilde{L}_0-\frac{c_R}{24}=\frac{H+J}{2}\;.
\end{align}
Note that in Euclidean signature the modular parameter $\t$ becomes
complex, and $\mu$ is purely imaginary. Furthermore, from the knowledge of the partition function, the zero modes of Virasoro generators can be simply obtained as:
\begin{equation}
\label{eq:inv_Vir_modes}
h_L=\frac{\rm i}{2\,\pi}\frac{\partial S_E}{\partial \t}\;,\quad h_R=\frac{-\rm i}{2\,\pi}\frac{\partial S_E}{\partial \bar\t}\;.
\end{equation}
From these generic considerations it is quite easy to obtain the
thermal partition function for ${\rm AdS}_3$.  Since global ${\rm AdS}_3$
corresponds to the NS-NS vacuum for which $L_0=\tilde{L}_0=0$, 
we get:
\begin{equation}
S_E=\frac{{\rm i}\,\pi}{12}\,(c_L\,\t-c_R\,\bar\t)\,.
\end{equation}
This is the result for the local part of the partition functions,
where non-local contributions given by excitations of massless 
particles can be neglected in the low temperature (large $\b$) limit.
By using the modular transformation $\t\rightarrow -\tfrac1{\t}$, one
gets the partition function for BTZ black holes:
\begin{equation}
S_{\rm E, BTZ}(\t,\bar\t)=S_E(-\tfrac1{\t},-\tfrac1{\bar \t})=-\frac{{\rm i}\,\pi}{12}\,\Big(\frac{c_L}{\t}-\frac{c_R}{\bar\t}\Big)\;,
\end{equation}
and, from \eqref{eq:inv_Vir_modes}, $h_L=-\frac{c_L}{24\,\t^2}$ and
$h_R=-\frac{c_R}{24\,\bar\t^2}$. Note now that the Helmholtz free energy corresponds to the Euclidean action, 
$S_E=\b\,H+\mu\,J-{\cS}$, so it is easy to get the Cardy formula:
\begin{equation}
\label{eq:btz_entroy}
\cS_{\rm BTZ}=2\,\pi\,\Big(\sqrt{\frac{c_L\,h_L}{6}}+\sqrt{\frac{c_R\,h_R}{6}}\Big)\;,
\end{equation}
 which is the expected result and can be obtained without relying on AdS/CFT correspondence. Since the entropy depends 
on $c_{L,R}$ (with $h_{L,R}$ fixed by \eqref{eq:tau_h}), we would like
to compute the central charges directly in gravity description which, in our
context, includes higher derivative corrections. This can be 
done by exploiting the anomalies of the system; the gravity side
suffers from anomalies arising from Chern-Simons 
couplings, and so does the CFT theory. The important observation here
is that the conformal anomalies are proportional to 
the central charges. It is important to point out that we are not
aiming at canceling the anomalies of either theory. We want 
to obtain the anomalies on both the gravity and the CFT side, and by
comparison, a formula for the central charges of the 
CFT in terms of the charges and coefficients characterizing the theory of gravity, including the higher derivative sector.
 There are many different but equivalent approaches to this problem 
\cite{ Harvey:1998bx,Kraus:2005vz,Maldacena:1997de,Freed:1998tg}. Here
we outline the line of reasoning and the main 
results, referring to the previous references for a detailed discussion.\\
 
The five dimensional ${\rm AdS}_3\times {\rm S}^2$ supergravity background, which is of interest to us, arises as a near-horizon limit of extremal BTZ black holes. The theory arises
by reducing M-theory on a Calabi-Yau three-fold $X$. From
the perspective of M-theory, we wrap $Q_5$ M5-branes with unity charge (or
equivalently one M5-brane, with charge $Q_5$) 
over the four-cycles of a Calabi-Yau three-fold $X$, hence reducing
the six-dimensional world-volume of the M5-branes to 
a 1+1 dimensional world-sheet spanned by a string in 5-dimensions. At
low energies, this string is described by a 
chiral (4,0) supersymmetric ${\rm CFT}_2$. 
The anomalous gravitational Chern-Simons term in the five dimensional
theory arises from the dimensional reduction of the 
M-theory coupling between the three-form potential $C_3$ and an eight-form term, proportional to the Riemann curvature to 
the fourth power. Each M5-brane is magnetically charged under the
three-form potential with unity charge. Now if we 
call $\{\theta^A\}$\footnote{The index A runs from 1 to $b^2=b_4$,
  where $b$ is the Betty number of the three-fold X.} an 
integral basis for the cohomolgy group $H^2(X,\mathbb{Z})$ of the
Calabi-Yau three-fold, and $\{\sigma_A\}$ the dual basis of 
the homology group $H_4(X,\mathbb{Z})$, then the three-form potential
reduces to a set of ${\rm U}(1)$ connection as 
$C_3= A^A_1\wedge \theta_A$. In the following, we will indicate  by $A_1$ the linear combination of abelian gauge fields  obtained from the element of $H^2$ dual to a smooth cycle
$P_0=P^A_0\,\sigma_A$ of $H_4$, i.e. 
\begin{equation}
A_1=a_{A}\,A_1^A \qquad a_A \cdot P_0^B=\delta_A^B\,. \nonumber
\end{equation}
By wrapping the $Q_5$ five-branes on the smooth cycles
$P_0$, one obtains a string in five-dimensional space, 
charged under the ${\rm U}(1)$ connections with charges $P^A_0$. In
our case we are wrapping $p^A$ unity charge 
M5-branes over the $A$-th four-cycle of ${\rm CY}_3$.\\

We will come back shortly to the microscopic stringy picture, but
first, it is instructive to understand the 
supergravity picture. We are considering a smooth near-horizon
configuration $\rm AdS_3\times S^2\times X$ of a higher derivative 
theory including the anomalous Chern-Simons couplings
\begin{equation}
\label{eq:CSs}
S_{anml}=\frac{1}{48}\,c_2\cdot P_0\int_{M_5} A_1\wedge p_1 \,,
\end{equation} 
 where $p_1=-\frac{1}{2(2\,\pi)^2}\,{\rm Tr}(R\wedge R)$ is the first
 Pontryagin class of the 2-form curvature $R$. Note that in
 this low-energy limit, the branes are interpreted as
 fluxes. Specifically, the magnetic charges of this solution are given by:
 \begin{equation}
 \label{eq:magn_charges}
 P_0^A=-\frac{1}{2\,\pi}\int_{S^2}\,F^A\;,\qquad Q_5=-\frac{1}{2\,\pi}\int_{S^2} F_1 \,,
 \end{equation}
 where $F^A= d A^A$ and $F_1= d A_1$ and we are assuming that the field strength has components only on the two-sphere. These are the same charges of the string configuration.
 Now the coupling \eqref{eq:CSs} suffers from two anomalies connected
 to diffeomorphism transformations on the tangent bundle and the normal bundle. 
Practically speaking, these anomalies correspond respectively  to the
anomalous transformations mapping the boundary onto the 
boundary and acting on the vectors normal to the boundary. In the
stringy perspective, the former corresponds to the tangent 
bundle anomaly on the string world-sheet, and is connected to the
gravitational anomalies of the ${\rm CFT}_2$. Specifically, if one
 uses Christoffel connections to define the curvature $R$, then
 the anomaly in the gravity side corresponds to the 
non-conservation of the stress-energy tensor. If instead, the spin
connection is used, then the anomaly corresponds to the
 antisymmetric part of the stress-energy tensor. 
 The normal bundle anomaly is interpreted as anomalous Lorentz transformation
 acting on the directions normal to the string world-sheet. 
Since we are in five dimensions, this group is ${\rm SO}(3)$, and is
associated to the R-symmetry group ${\rm SU}(2)_L$ acting on the 
left-mover degrees of freedom of the (4,0) CFT.\\ 
 Let us start analyzing the tangent bundle anomaly. On the supergravity
 side, under an infinitesimal diffeomorphism transformation 
of the form $x^\mu\rightarrow x^\mu-\xi^\mu$ \footnote{Note that
  diffeomorphism anomalies are possible only in even dimensions, 
so in this context, these anomalies arise as boundary symmetries that
are broken in the quantum theory}, the transformations 
rules are as follows (see \cite{Ginsparg:1985qn,Harvey:2005it} for a pedagogical discussion):
 \begin{equation}
\delta_\xi g_{\mu\nu}=2\,\xi_{(\mu\nu)}\;,\quad \delta_\xi\Gamma={\rm d}\xi+[\Gamma,\xi]\;,\quad \delta_\xi R=[R,\xi]\;,
 \end{equation}
 where $\xi_{\mu\nu}=\frac{\partial \xi_\mu}{\partial x^\nu}$. Now the anomalous term \eqref{eq:CSs} can be written as:
 \begin{equation}
S_{anml}=-\frac{1}{(2\,\pi)^2}\,\frac{1}{96}\,c_2\cdot P_0\int_{M_5}
A\wedge {\rm Tr}(R\wedge R)= \frac{1}{(2\,\pi)^2}\,
\frac{1}{96}\,c_2\cdot P_0\int_{M_5} F_1\wedge {\rm Tr}(\Omega_3)\;,
 \end{equation}
where, we defined $\Omega_3={\rm Tr}(\Gamma\,{\rm d}
\Gamma +\tfrac23\,\Gamma^3)$. The above equality represents the first step of the descent procedure,
i.e. ${\rm Tr}(R\wedge R)={\rm d}\Omega_3$; the integrand proportional to the gauge field strength can be made well-defined by partially integrating it, as the linear combination
of gauge fields $A_1$ is ill-defined in the presence of 
magnetic charges.  Next, we exploit the 
fact that the field strength $F_1$ has components only over the sphere, to work out the integral and get:
 \begin{equation}
 S_{anml}=-\frac{1}{(2\,\pi)}\,\frac{1}{96}\,c_2\cdot p\int_{AdS_3} {\rm Tr}(\Omega_3) \,.
 \end{equation}
Now, we just need to apply the second step of the descent
procedure, i.e. we consider a symmetry variation and write 
it in terms of an exact form. Since $\delta_\xi \Omega_3={\rm d}\,{\rm Tr}(v\,{\rm d}\Gamma)$, we have:
\begin{equation}
 \delta_\xi S_{anml}=-\frac{1}{(2\,\pi)}\,\frac{1}{96}\,c_2\cdot p\int_{\partial AdS_3} {\rm Tr}(v\,{\rm d}\Gamma) \,.
\end{equation}
This shows that the diffeomorphism invariance of the bulk theory is
preserved while the invariance of the boundary theory is 
anomalous. But this anomaly can be equaled to the
gravitational anomaly of the ${\rm CFT}_2$, which can also be computed through the descent procedure, from the four form 
$I_4=-\tfrac{(c_L-c_R)p_1}{24}$. We finally have:
\begin{equation}
\delta_\xi S_{CFT}=\frac{c_R-c_L}{48\,(2\pi)}\int_{\partial AdS_3}\,{\rm Tr}(v\,d\Gamma) \,.
\end{equation}
 By simple comparison we find the identity:
 \begin{equation}
 \label{eq:first_constr}
 c_L-c_R=\tfrac12\,c_2\cdot p \,.
 \end{equation}
This is not enough to constrain both the central charges, but
we do have another anomaly to take into account, the normal 
bundle anomaly, which corresponds to R-symmetry anomaly in the
$(4,0)$ CFT side. In fact, since our theory is chiral, it will be anomalous only in the supersymmetric sector (leftmovers in our
conventions). The analysis for the normal bundle 
anomaly turns out to be more involved than its tangent bundle
counterpart. We note here that, since the gravity theory we 
are considering is supersymmetric, the two anomalies must
necessarily fit in the same multiplet, i.e. be connected by 
supersymmetry transformations. Furthermore, as was pointed out in
\cite{Kraus:2005zm} the Chern-Simons term treats
 the left and right central charges oppositely (we refer also to
 \cite{Castro:2007sd}, where a crucial result for the analysis of 
\cite{Kraus:2005zm} is found). Here we follow the reasoning of
\cite{Harvey:1998bx}; as we said before, the normal bundle 
anomaly for a string in 5 dimensions is connected to the subgroup
${\rm SO}(3)$ of the Lorentz group, which leaves the string 
invariant (hence acting on its transverse directions). Specifically, the
left-movers superalgebra is an ${\rm SO}(3)$ Kac-Moody 
algebra with level $k$, which is connected, by the AdS/CFT
correspondence, to the coefficients of the Chern-Simons 
terms. Furthermore, the superconformal algebra constrains the (left,
in this case) central charge to be equal to $c_L=6\,k$. All 
that is left is for us to compute the level $k$. We refrain from
giving the details of the calculation; the interested reader should 
carefully read  \cite{Harvey:1998bx}. The final result, which also
includes the contribution from the gauge Chern-Simons 
coupling $A\wedge F\wedge F$, reads:
\begin{equation}
\label{eq:second_constr}
k=D_0+\tfrac1{12}c_2\cdot p \,,
\end{equation}
where $D_0=\tfrac16\, D_{ABC}\,p^A\,p^B\,p^C$ and $D_{ABC}$ are the intersection numbers of $X$.
Using \eqref{eq:first_constr} and \eqref{eq:second_constr} we finally find the values of the central charges
\begin{equation}
\label{eq:c_RL}
c_L=D_{ABC}\,p^A\,p^B\,p^C+\tfrac12\,c_2\cdot p\;,\qquad c_R=D_{ABC}\,p^A\,p^B\,p^C+c_2\cdot p\;.
\end{equation}
The result \eqref{eq:c_RL} depends only on the Chern-Simons
terms of the five (or three) dimensional theory, which 
are anomalous and contribute to both the left (BPS) and the right
(non-BPS) central charges of the boundary CFT theory. Hence, since no other anomalous terms should exists in 5D, the above results should be correct at any order in the effective pertubative expansion of the theory.\\
Now, to obtain an explicit result for the entropy, it is just a matter of finding the value of $h_{L,R}$ for the classes of (non-)BPS ${\rm AdS}_3\times {\rm S}^2$ background. 

\subsection{5D vs 4D Theories and Solutions}
We are interested in computing the subleading corrections to the entropy of
 four-dimensional black hole solutions in $\cN=2$ 
supergravity, with the near-horizon geometry ${\rm AdS}_2\times {\rm S}^2$. Hence we
need to justify the connection between the results obtained for five dimensional black holes in the previous section and the results we aim to obtain.
First of all, as already emphasized, the statistical entropy
computation of the five dimensional black holes is
sensitive only to the anomalous terms, which do not exist in 
the four dimensional theory. Secondly, the BTZ background analyzed is
obtained as the compatification of an M-theory solution 
over a Calabi-Yau three-fold and hence as the (non-)BPS
solution of a supersymmetric theory. On the other hand, the 
macroscopic methods available to compute the entropy of supergravity
black holes are based on the knowledge of the full 
Lagrangian, and no supersymmetry is required. If we wish to exploit
the five-dimensional entropy results for the 
four-dimensional backgrounds of interest in this work, we need to
find the connection between the two theories and the two corresponding
backgrounds. The latter is quite obvious; the five dimensional
${\rm AdS}_3\times {\rm S}^2$ background is related to the four dimensional
${\rm AdS}_2\times {\rm S}^2$ background via a simple circle
reduction. Analogously, the five dimensional supergravity theory, with Weyl squared higher-derivative corrections \cite{Hanaki:2006pj} can be reduced over a
circle and the reduction procedure has been worked out in great detail in
\cite{Banerjee:2011ts}. The results showed 
the presence of some new higher-derivative terms, belonging to no known four 
derivatives invariant in four dimensions.
In retrospect, it is very easy to guess that the five-dimensional
gravitational Chern-Simons term $A_5 \wedge R_5\wedge R_5$, 
with $R_5$ being the Riemann tensor in 5D, will reduce to terms proportional to
$X^0_4\,R_4\wedge R_4$ in 4D, which is squared in the 
curvatures, but not proportional to the squared Weyl tensor.\\

Shortly after, these curvature squared terms were found in the purely
 four-dimensional context, as part of the bosonic sector 
of a new class of higher derivative invariants, TLog invariants, built out of a
non-linear chiral multiplet \cite{Butter:2013lta}. Contrary to the 
Weyl squared invariant, which depends only on superconformal gravity
fields, the new class of invariants depends explicitly on 
matter (or gauge) multiplets. Furthermore, by combining the TLog invariant, for a constant compensating multiplet 
configuration, and the Weyl
squared Lagrangian , one obtains the Gauss-Bonnet density, which is a
topological invariant of four dimensional gravity theory.\\
We want to stress that it is 
of utmost importance for the four-dimensional theory under
consideration to be the exact reduction of the supersymmetrization 
of the five-dimensional gravitational Chern-Simons terms. Even though the entropy computation depends only on the two Chern-Simons terms, 
supersymmetry of the full action plays a crucial role. This is the
result presented in \eqref{eq:full_prepot}. We refer the reader to 
the last section of \cite{Butter:2014iwa} for further details.\\

At this point, it should be clear that the results
obtained in the previous section can also be used in the four-dimensional context without any 
modification. This would require only the identification of
the reduced background and the reduced theories correctly \cite{Butter:2014iwa,Banerjee:2011ts}.
We are now ready to present the details of the entropy computation for
a class of BPS and non-BPS black hole solutions, and show
 the exact agreement with the entropy results \eqref{eq:btz_entroy} and \eqref{eq:c_RL}.
 
\subsection{Entropy of (non-)BPS Black Holes}
The classes of black holes of interest to this work are obtained as solutions of superconformal gravity coupled to three vector multiplets, labelled by the indices A, B and C. However the number of vector multiplets in our solution is by no means constrained, i.e. the results we obtain in this section can be generalized to any number of vector multiplets. Doing so would slightly change the physical interpretation, as then we would be considering an effective low energy description of M-theory compactified over a Calabi-Yau manifold \cite{Maldacena:1997de} (see also \cite{Sahoo:2006rp}).\\

We start with BPS black hole solutions, which were first analyzed in \cite{LopesCardoso:1998tkj,LopesCardoso:1999cv}. On imposing full supersymmetry of the background, sufficient conditions extremizing the entropy function can be found. They are:
\begin{equation}
\label{eq:BPS_attractors}
v_1=v_2=\frac{16}{z\,\bar z}\;,\qquad\cF^{-\,I}_{01}=e^I- {\rm i}\, \frac{v_1}{v_2}\,p^I-\tfrac12\,\bar X^I\,v_1\,z=0\;,\qquad \frac{\bar F_I}{\bar z}-\frac{F_I}{z}=-\frac{\rm i}{4}\,q_I\;.
\end{equation}
The first two conditions are sufficient to make the whole TLog invariant collapse to zero \cite{Butter:2014iwa}. The only non-vanishing terms in the higher derivative sector come from the highest component of the Weyl squared invariant.
The conditions above are steadily solved by :
\begin{equation}
e^I=4\Big(\frac{\bar X^I}{\bar z}+\frac{X^I}{z}\Big)\;,\qquad p^I=4\, {\rm i}\, \Big(\frac{\bar X^I}{\bar z}-\frac{X^I}{z}\Big) \,,
\end{equation}
from which we derive a general formula for the entropy, given by:
\begin{equation}
S_{\rm BH}=2\,\pi\Big[-\tfrac12 e^I q_I-16\, {\rm i} \Big(\frac{F}{z^2}-\frac{\bar F}{\bar z^2}\Big)\Big] \,.
\end{equation}
Note that the entropy still depends on the electric fields $e^I$, the vector multiplet scalars $X^I$ and the T-tensor component $z$. To fix those we exploit the invariance of the equations of motion, derived from the Lagrangian \eqref{eq:Lagrangian_prepot} with the prepotential \eqref{eq:prepotential}, under ${\rm SO}(2,2)={\rm SL}(2,\mathbb{Z})\times {\rm SL}(2,\mathbb{Z})$ T-duality. This means that we can choose a representative set of charges and electric fields, to simplify the computation. In this case, we can choose as representative satisfying the $\mathbb{Z}_2$ symmetry that the four dimensional theory inherits from M-theory, i.e.:
\begin{equation}
p^i\rightarrow p^A\;,\quad p^0\rightarrow -p^0\;,\quad e^A\rightarrow -e^A\;,\quad e^0\rightarrow e^0\;,\quad X^A\rightarrow- \bar X^A\;,\quad X^0\rightarrow \bar X^0\;,\quad z\rightarrow \bar z\;.
\end{equation} 
Thus we can choose a gauge (or exploit $\mathbb{Z}_2$ invariance) to fix the charges and the moduli as:
\begin{equation}
p^0=0\;,\qquad\qquad e^A=0\;,\qquad\qquad X^A= {\rm i}\, y^A\;,\qquad\qquad X^0=y^0\;,
\end{equation} 
with $z, y^i \in \mathbb{R}$. Note that this also implies $q^A=0$.\\
For the higher derivative theory \eqref{eq:Lagrangian_prepot}, the supersymmetric attractors conditions \eqref{eq:BPS_attractors} are satisfied by:
\begin{align}
\label{eq:BPS_near_horizon}
w&=1\;,\qquad\qquad  v_1=v_2=16\;,\qquad\qquad x^A=\tfrac18 \,{\rm i}\,p^A\;, \qquad\qquad e^A=0,  \nonumber \\
x^0&=\tfrac18\,\sqrt{\frac{d_{ABC}\,p^A\,p^B\,p^C+256\,d_A\,p^A}{-q_0}}\;,\quad  
e^0=\sqrt{\frac{d_{ABC}\,p^A\,p^B\,p^C+256\,d_A\,p^A}{-q_0}}\;,
\end{align}
for $q_0<0$ and $d_{ABC}\,p^A\,p^B\,p^C+256\,d_A\,p^A>0$. Note at this point that the KK-charge $q_0$ in four dimensions is identified with the angular momentum $J$ of the ${\rm BTZ}\times S^2$ solution in 5D, for which, in the extremal limit, $H=|J|$. The class of BPS solutions analyzed satisfies $H=-J$, so from eq. \eqref{eq:tau_h} we get $h_L=0$ and $h_R=-J=-q_0>0$. 
 This indeed leads to the result of \cite{LopesCardoso:1998tkj}, i.e.,
\begin{equation}
S_{\rm BH}=2\,\pi\,\sqrt{-q_0(d_{ABC}\,p^A\,p^B\,p^C+256\,d_A\,p^A)}\;.
\end{equation}
We point out that the constraints imposed by the BPS attractors are very stringent and allow for an analytical solution for all the fields, even when the higher derivative sector is considered. That is why the final result is already in the form expected from the analogous statistical computation, i.e. the Cardy formula. As we stressed before, the situation will not be quite as simple for the non-BPS solutions, as a closed form analytical solution will, in general, be missing. \\

Next, we procede with the computation of the entropy of non-BPS solutions, studied in \cite{Goldstein:2005hq,Tripathy:2005qp}. This class of solutions exists for the two derivative theory, $d_A=0$, and has the same charge configuration as the BPS solution above except for one difference, residing in a simple sign change for the electric field $e^0$, consistent with the $\mathbb{Z}_2$ duality symmetry of the two derivative action. The auxiliary field $z$ cannot be fixed by this procedure and we need to solve for it via the extremization equations. The results hence read:
\begin{align}
\label{eq:nonBPS_near_horizon}
w&=\tfrac{1}{2}\;,\qquad v_1=v_2=16\;,\qquad\qquad  x^A=\tfrac18 \,{\rm i}\,p^A\;,
                                                \qquad\qquad   e^A=0\;,
\nonumber\\
x^0&=\tfrac18\,\sqrt{\frac{d_{ABC}\,p^A\,p^B\,p^C}{q_0}}\;, \qquad
e^0=-\sqrt{\frac{d_{ABC}\,p^A\,p^B\,p^C}{q_0}}\;,
\end{align}
for $q_0>0$ and $d_{ABC}\,p^A\,p^B\,p^C>0$. The leading order entropy reads:
\begin{equation}
S_{\rm BH}=2\,\pi\,\sqrt{q_0(d_{ABC}\,p^A\,p^B\,p^C)}\;.
\end{equation}
To compute the first order corrections in $d_A$ we just need to plug
this leading order solution into the full entropy function, 
as explained before. Note that, if we would have used Wald
formalism, we would have needed the full near-horizon solution 
for the parameters, so we could not have used the solution presented
here, which was found from an $\cN=2$ two derivative 
theory \cite{Tripathy:2005qp}. Note again that, since we are flipping
the sign of $q_0=J$, the condition of 
extremality $H=J$ forces $h_R=0$ and $h_L=J=q_0>0$. So the first
subleading order correction to the entropy of this class 
of black holes is now given by:
\begin{equation}
S_{\rm BH}=2\,\pi\,\sqrt{q_0(d_{ABC}\,p^A\,p^B\,p^C)}\Big(1+\frac{64\,d_A\,p^A}{d_{ABC}\,p^A\,p^B\,p^C}\Big)\;,
\end{equation}
which corresponds, as we announced before, to the correct expansion up to the first order of Cardy formula (eq. \eqref{eq:btz_entroy}). Hence we have resolved the puzzle concerning the macroscopic entropy of non-BPS black holes in 5D/4D. 

\section{Results and Conclusions}
\label{sec:end}
In this work we have resolved an entropy puzzle for four dimensional
non-BPS black holes of $\cN=2$ supergravity. We find perfect
agreement, in the thermodynamic limit of large charges, between our macroscopic entropy results and the expected statistical 
entropy computed in \cite{Kraus:2005vz}, even for
non-supersymmetric black holes. The key to obtain this
result is the knowledge of the exact connection between 
the five and four-dimensional theories. Specifically, the
five-dimensional result is obtained from the anomaly calculations in 5D, 
where the anomalous terms contributing to the entropy are the
Chern-Simons interactions. To match this result in 4D, the need for the 
full reduction of the supersymmetric gravitational Chern-Simons term
was imperative \cite{Banerjee:2011ts}, and the relevant results in \cite{Butter:2013lta,Butter:2014iwa} have been used extensively in this work 
Based on the results of
\cite{Harvey:1998bx,Kraus:2005vz,Kraus:2005zm}, it would surely seem
that no other higher order 
invariant would contribute to the entropy of (non-)BPS black holes in four
dimensions, simply because the statistical 5D results
that hinge on anomalies, cannot be corrected further, as no more
anomalous terms are supposed to play a role at any order 
in the perturbative expansion of the 5D theory. The computation is
very solid and little doubt exists about its validity. In fact, even though in the previous sections we restricted ourselves to first order corrections to the entropy\footnote{This is, in principle, the maximal accuracy we can reach by considering the lowest order higher derivative action.}, one can push the computation forward, by iteratively computing the higher order corrections to the near-horizon parameter and entropy. It turns out that the series expansion of entropy coincides exactly with the series expansion of the square root, i.e.
\begin{align}
S_{\rm BH}&=2\,\pi\,\sqrt{q_0(d_{ABC}\,p^A\,p^B\,p^C)}\Big(1+64\,u-2048 u^2+131072 u^3-10485760 u^4 + 939524096 u^5+\dots\Big)
\nonumber\\
&=2\,\pi\,\sqrt{q_0(d_{ABC}\,p^A\,p^B\,p^C+128 d_A\,p^A)}\;,
\end{align}
with $u=\frac{d_A\,p^A}{d_{ABC}\,p^A\,p^B\,p^C}$ being the smallness parameter. Hence, as consistency dictates, the microscopic entropy computed from anomalies matches exactly the macroscopic entropy obtained from the knowledge of the full effective action containing the anomalous terms. Of course, this proves that, for the class on \emph{large} black holes considered herein, the anomaly result is reliable also in the non-supersymmetric sector.
 Nevertheless, there exist indications that the anomaly results might not work for \emph{every} class of black holes. In fact, although the supersymmetric gravitational Chern-Simons term is the most analyzed higher derivative invariant in 5D,
other invariants were recently constructed \cite{Ozkan:2013nwa}. These invariants also include the Gauss-Bonnet
density in 5D, which was analyzed in the context of 
black hole entropy computations many years before
\cite{Cvitan:2007hu}. As it turns out, small black holes seem to
escape 
the description in terms of the Chern-Simons terms, and instead
receive corrections from the disconnected Gauss-Bonnet 
sector. This means that there might be some hidden caveat in the
reasoning of \cite{ Kraus:2005vz} which is evidently 
contradicted in the presence of small black holes.
If the 5D theory possesses many sectors, each of which
is connected to special classes of black holes, then it 
is obvious to assume that something similar would also happen in
4D theory, i.e. other higher derivative invariants, of the 
first order in the perturbative expansion in $d_A$ might exist in
$\cN=2$ theory. Furthermore, it would be important to 
understand how exactly the anomaly procedure that led to the correct
entropy results for the classes of black holes analyzed, 
fails to capture information about different classes, and its exact range of applicability.
We plan on addressing these issues in the near-future.

\section*{Acknowledgements}
We are thankful to Daniel Butter, Suvankar Dutta, Sameer Murthy, Bernard de Wit,
Dileep Jatkar, Bindusar Sahoo and Ashoke Sen for discussions at
various stages of this work. Ivano Lodato thanks the organizers of NSM
Mohali, where this work was first presented. NB and IL are partly
supported by a DST Ramanujan Grant. Finally, we are thankful to the
people of India for their generous support to basic science. 

\appendix

\section{Conventions and Useful Identities}
\label{app.A}
\renewcommand{\theequation}{A.\arabic{equation}}
We use the Pauli-K\"all\'{e}n convention. Spacetime indices are denoted $\mu,\nu,\ldots$, Lorentz indices are denoted
$a,b,\ldots$, and $\mathrm{SU}(2)$ indices are denoted $i,j,\ldots$.
The Lorentz metric is $\eta_{ab} = \textrm{diag}(-1,1,1,1)$ and the
anti-symmetric tensor $\veps_{abcd}$ is imaginary, with $\veps_{0123} = -{\rm i} = -\, \veps^{0123}$.\\
An anti-symmetric two-form $F_{ab}$ satisfies the following identities:
\begin{gather}
F_{ab}^\pm = \tfrac{1}{2} (F_{ab} \pm \tilde F_{ab})~, \qquad
\tilde F_{ab} = \tfrac{1}{2} \veps_{abcd} F^{cd}~, \qquad
\tilde F_{ab}^\pm = \pm F_{ab}^\pm~. \label{eq:Fselfdual}
\end{gather}
We always apply symmetrization and anti-symmetrization with unit strength,
so that $F_{[ab]} = F_{ab}$ and $F_{(\alpha \beta)} = F_{\alpha \beta}$.
Furthermore, the following useful identities for products of (anti-)selfdual tensors are noted,
\begin{align}
  \label{eq:dual-tensor-products}
  G^\pm_{[a[c}\, H_{d]b]}^\pm =&\, \pm\ft18
  G^\pm_{ef} \,H^{\pm ef} \,\varepsilon_{abcd} -\ft14( G^\pm_{ab}\,
  H^\pm_{cd} +G^\pm_{cd}\, H^\pm_{ab}) \,,\nonumber \\
  G^{\pm}_{ab} \,H^{\mp cd} + G^{\pm cd} \, H^{\mp}_{ab} =&\, 4 \delta^{[c}_{[a}
  G^{\pm}_{b]e} \, H^{\mp d]e} \,,\nonumber \\
  \ft12 \varepsilon^{abcd} \,G^{\pm}_{[c}{}^e \, H^\pm_{d]e} =&\, \pm
  G^{\pm [a}{}_{e} \, H^{\pm b]e}\,,\nonumber \\
  G^{\pm ac}\,H^\pm_c{}^b + G^{\pm bc}\,H^\pm_c{}^a=&\, -\ft12
  \eta^{ab}\, G^{\pm cd}\,H^\pm_{cd} \,,\nonumber \\
  G^{\pm ac}\,H^\mp_c{}^b =&\, G^{\pm bc}\,H^\mp_c{}^a\,,\nonumber\\
  G^{\pm ab}\,H^\mp_{ab} =&\,0 \,.
\end{align}
Finally, we remind the reader that $\rm SU(2)$ indices are swapped by
complex conjugation, $(T_{abij})^* = T_{ab}{}^{ij}$, and we make use
of the invariant $\rm SU(2)$ tensor $\veps^{ij}$ and $\veps_{ij}$
defined as $\veps^{\1\2} = \veps_{\1\2} = 1$ with $\veps^{ij}
\veps_{kj} = \delta^i_k$.

\section{Superconformal Mutiplets -  (Covariant) Weyl, Chiral, Vector multiplets}
\label{app.B}
\renewcommand{\theequation}{B.\arabic{equation}}
 Recall that the superconformal algebra
comprises the generators of the general-coordinate, local Lorentz,
dilatation, special conformal, chiral $\mathrm{SU}(2)$ and
$\mathrm{U}(1)$, supersymmetry (Q) and special supersymmetry (S)
transformations. 
The gauge fields associated with general-coordinate
transformations ($e_\mu{}^a$), dilatations ($b_\mu$), R-symmetry
($\mathcal{V}_\mu{}^i{}_j$ and $A_\mu$) and Q-supersymmetry
($\psi_\mu{}^i$) are independent fields.  The remaining gauge fields
associated with the Lorentz ($\omega_\mu{}^{ab}$), special conformal
($f_\mu{}^a$) and S-supersymmetry transformations ($\phi_\mu{}^i$) are
composite objects.  The
multiplet also contains three other fields: a Majorana spinor doublet
$\chi^i$, a scalar $D$, and a selfdual Lorentz tensor $T_{abij}$,
which is anti-symmetric in $[ab]$ and $[ij]$. The Weyl and chiral
weights have been collected in table \ref{table:weyl}.
%
Under Q-supersymmetry, S-supersymmetry and special conformal
transformations, the Weyl multiplet fields transform
as
\begin{eqnarray}
  \label{eq:weyl-multiplet}
  \delta e_\mu{}^a & =& \bar{\epsilon}^i \, \gamma^a \psi_{ \mu i} +
  \bar{\epsilon}_i \, \gamma^a \psi_{ \mu}{}^i \, , \nonumber\\
  \delta \psi_{\mu}{}^{i} & =& 2 \,\mathcal{D}_\mu \epsilon^i - \tfrac{1}{8}
  T_{ab}{}^{ij} \gamma^{ab}\gamma_\mu \epsilon_j - \gamma_\mu \eta^i
  \, \nonumber \\
  \delta b_\mu & =& \tfrac{1}{2} \bar{\epsilon}^i \phi_{\mu i} -
  \tfrac{3}{4} \bar{\epsilon}^i \gamma_\mu \chi_i - \tfrac{1}{2}
  \bar{\eta}^i \psi_{\mu i} + \mbox{h.c.} + \Lambda^a_{\rm K} e_{\mu a} \, ,
  \nonumber \\
  \delta A_{\mu} & =& \tfrac{1}{2} \mathrm{i} \bar{\epsilon}^i \phi_{\mu i} +
  \tfrac{3}{4} \mathrm{i} \bar{\epsilon}^i \gamma_\mu \, \chi_i +
  \tfrac{1}{2} \mathrm{i}
  \bar{\eta}^i \psi_{\mu i} + \mbox{h.c.} \, , \nonumber\\
  \delta \mathcal{V}_\mu{}^{i}{}_j &=& 2\, \bar{\epsilon}_j
  \phi_\mu{}^i - 3
  \bar{\epsilon}_j \gamma_\mu \, \chi^i + 2 \bar{\eta}_j \, \psi_{\mu}{}^i
  - (\mbox{h.c. ; traceless}) \, , \nonumber \\
  \delta T_{ab}{}^{ij} &=& 8 \,\bar{\epsilon}^{[i} R(Q)_{ab}{}^{j]} \,
  , \nonumber \\
  \delta \chi^i & =& - \tfrac{1}{12} \gamma^{ab} \, \Slash{D} T_{ab}{}^{ij}
  \, \epsilon_j + \tfrac{1}{6} R(\mathcal{V})_{\mu\nu}{}^i{}_j
  \gamma^{\mu\nu} \epsilon^j -
  \tfrac{1}{3} \mathrm{i} R_{\mu\nu}(A) \gamma^{\mu\nu} \epsilon^i + D
  \epsilon^i +
  \tfrac{1}{12} \gamma_{ab} T^{ab ij} \eta_j \, , \nonumber \\
  \delta D & =& \bar{\epsilon}^i \,  \Slash{D} \chi_i +
  \bar{\epsilon}_i \,\Slash{D}\chi^i \, .
\end{eqnarray}
Here $\epsilon^i$ and $\epsilon_i$ denote the spinorial parameters of
Q-supersymmetry, $\eta^i$ and $\eta_i$ those of S-supersymmetry, and
$\Lambda_{\rm K}{}^a$ is the transformation parameter for special conformal
boosts.  The full superconformally covariant derivative is denoted by
$D_\mu$, while $\mathcal{D}_\mu$ denotes a covariant derivative with
respect to Lorentz, dilatation, chiral $\mathrm{U}(1)$ and
$\mathrm{SU}(2)$ transformations,
\begin{equation}
  \label{eq:D-epslon}
  \mathcal{D}_{\mu} \epsilon^i = \big(\partial_\mu - \tfrac{1}{4}
    \omega_\mu{}^{cd} \, \gamma_{cd} + \tfrac1{2} \, b_\mu +
    \tfrac{1}{2}\mathrm{i} \, A_\mu  \big) \epsilon^i + \tfrac1{2} \,
  \mathcal{V}_{\mu}{}^i{}_j \, \epsilon^j  \,.
\end{equation}
\begin{table}[t]
\begin{tabular*}{\textwidth}{@{\extracolsep{\fill}}
    |c||cccccccc|ccc||ccc| }
\hline
\noalign{\smallskip}
 & &\multicolumn{9}{c}{Weyl Multiplet Components} & &
 \multicolumn{2}{c}{Parameters} & \\[1mm]  \hline \hline 
 \noalign{\smallskip}
 field & $e_\mu{}^{a}$ & $\psi_\mu{}^i$ & $b_\mu$ & $A_\mu$ &
 $\mathcal{V}_\mu{}^i{}_j$ & $T_{ab}{}^{ij} $ &
 $ \chi^i $ & $D$ & $\omega_\mu^{ab}$ & $f_\mu{}^a$ & $\phi_\mu{}^i$ &
 $\epsilon^i$ & $\eta^i$
 & \\[1mm] \hline\noalign{\smallskip}
$w$  & $-1$ & $-\tfrac12 $ & 0 &  0 & 0 & 1 & $\tfrac{3}{2}$ & 2 & 0 &
1 & $\tfrac12 $ & $ -\tfrac12 $  & $ \tfrac12  $ & \\[1mm] \hline\noalign{\smallskip}
$c$  & $0$ & $-\tfrac12 $ & 0 &  0 & 0 & $-1$ & $-\tfrac{1}{2}$ & 0 &
0 & 0 & $-\tfrac12 $ & $ -\tfrac12 $  & $ -\tfrac12  $ & \\[1mm] \hline\noalign{\smallskip}
 $\gamma_5$   &  & $\;+$ &   &    &   &   & $\;+$ &  &  &  & $\;-$ & 
 $ \;+ $  & $\;-$ & \\ \hline
\end{tabular*}
\vskip 2mm
\renewcommand{\baselinestretch}{1}
\parbox[c]{\textwidth}{\caption{\label{table:weyl}{\footnotesize
Weyl and chiral weights ($w$ and $c$) and fermion
chirality $(\gamma_5)$ of the Weyl multiplet component fields and the
supersymmetry transformation parameters.}}}
\end{table}

The covariant curvatures are given by
\begin{align}
  \label{eq:curvatures}
  R(P)_{\mu \nu}{}^a  = & \, 2 \, \partial_{[\mu} \, e_{\nu]}{}^a + 2 \,
  b_{[\mu} \, e_{\nu]}{}^a -2 \, \omega_{[\mu}{}^{ab} \, e_{\nu]b} -
  \tfrac1{2} ( \bar\psi_{[\mu}{}^i \gamma^a \psi_{\nu]i} +
  \mbox{h.c.} ) \, , \nonumber\\[.2ex]
  R(Q)_{\mu \nu}{}^i = & \, 2 \, \mathcal{D}_{[\mu} \psi_{\nu]}{}^i -
  \gamma_{[\mu}   \phi_{\nu]}{}^i - \tfrac{1}{8} \, T^{abij} \,
  \gamma_{ab} \, \gamma_{[\mu} \psi_{\nu]j} \, , \nonumber\\[.2ex]
  R(A)_{\mu \nu} = & \, 2 \, \partial_{[\mu} A_{\nu ]} - \mathrm{i}
  \left( \tfrac12
    \bar{\psi}_{[\mu}{}^i \phi_{\nu]i} + \tfrac{3}{4} \bar{\psi}_{[\mu}{}^i
    \gamma_{\nu ]} \chi_i - \mbox{h.c.} \right) \, , \nonumber\\[.2ex]
  R(\mathcal{V})_{\mu \nu}{}^i{}_j =& \, 2\, \partial_{[\mu}
  \mathcal{V}_{\nu]}{}^i{}_j +
  \mathcal{V}_{[\mu}{}^i{}_k \, \mathcal{V}_{\nu]}{}^k{}_j  +  2 (
    \bar{\psi}_{[\mu}{}^i \, \phi_{\nu]j} - \bar{\psi}_{[\mu j} \,
    \phi_{\nu]}{}^i )
  -3 ( \bar{\psi}_{[\mu}{}^i \gamma_{\nu]} \chi_j -
    \bar{\psi}_{[\mu j} \gamma_{\nu]} \chi^i ) \nonumber\\
& \, - \delta_j{}^i ( \bar{\psi}_{[\mu}{}^k \, \phi_{\nu]k} -
  \bar{\psi}_{[\mu k} \, \phi_{\nu]}{}^k )
  + \tfrac{3}{2}\delta_j{}^i (\bar{\psi}_{[\mu}{}^k \gamma_{\nu]}
  \chi_k - \bar{\psi}_{[\mu k} \gamma_{\nu]} \chi^k)  \, , \nonumber\\[.2ex]
  R(M)_{\mu \nu}{}^{ab} = & \,
  \, 2 \,\partial_{[\mu} \omega_{\nu]}{}^{ab} - 2\, \omega_{[\mu}{}^{ac}
  \omega_{\nu]c}{}^b
  - 4 f_{[\mu}{}^{[a} e_{\nu]}{}^{b]}
  + \tfrac12 (\bar{\psi}_{[\mu}{}^i \, \gamma^{ab} \,
  \phi_{\nu]i} + \mbox{h.c.} ) \nonumber\\
& \, + ( \tfrac14 \bar{\psi}_{\mu}{}^i   \,
  \psi_{\nu}{}^j  \, T^{ab}{}_{ij}
  - \tfrac{3}{4} \bar{\psi}_{[\mu}{}^i \, \gamma_{\nu]} \, \gamma^{ab}
  \chi_i
  - \bar{\psi}_{[\mu}{}^i \, \gamma_{\nu]} \,R(Q)^{ab}{}_i
  + \mbox{h.c.} ) \, , \nonumber\\[.2ex]
  R(D)_{\mu \nu} = & \,2\,\partial_{[\mu} b_{\nu]} - 2 f_{[\mu}{}^a
  e_{\nu]a}
  - \tfrac{1}{2} \bar{\psi}_{[\mu}{}^i \phi_{\nu]i} + \tfrac{3}{4}
    \bar{\psi}_{[\mu}{}^i \gamma_{\nu]} \chi_i
    - \tfrac{1}{2} \bar{\psi}_{[\mu i} \phi_{\nu]}{}^i + \tfrac{3}{4}
  \bar{\psi}_{[\mu i} \gamma_{\nu]} \chi^i \,,  \nonumber\\[.2ex]
  R(S)_{\mu\nu}{}^i  = \,&  2\,{\cal D}_{[\mu}\phi_{\nu]}{}^i
  -2 f_{[\mu}{}^a\gamma_a\psi_{\nu]}{}^i
  -\ft18 \Slash{D} T_{ab}{}^{ij}\gamma^{ab}\gamma_{[\mu} \psi_{\nu]\, j}
     -\tfrac32 \gamma_a\psi_{[\mu}{}^i\,\bar\psi_{\nu]}{}^j\gamma^a{\chi}_j
     \nonumber\\
     \,& +\ft14 R({\cal V})_{ab}{}^i{}_j\gamma^{ab}
     \gamma_{[\mu}\psi_{\nu]}{}^j
     +\ft12 \mathrm{i}
     R(A)_{ab}\gamma^{ab}\gamma_{[\mu}\psi_{\nu]}{}^i
     \,,\nonumber\\[.2ex]
     R(K)_{\mu\nu}{}^a = \,& 2 \,{\cal D}_{[\mu} f_{\nu]}{}^a
     -\ft14\big(\bar{\phi}_{[\mu}{}^i\gamma^a\phi_{\nu]i}
     +\bar{\phi}_{[\mu i} \gamma^a\phi_{\nu]}{}^i\big)  \nonumber\\
     &\,
     +\tfrac14\big(\bar{\psi}_{\mu }{}^iD_b T^{ba}{}_{ij}\psi_{\nu}{}^j
     -3\, e_{[\mu}{}^a\psi_{\nu]}{}^i\Slash{D}\chi_i +\ft32
     D\,\bar{\psi}_{[\mu}{}^i\gamma^a\psi_{\nu]j}
     -4\,\bar{\psi}_{[\mu}{}^i\gamma_{\nu]}D_b R(Q)^{ba}{}_i
     +\mbox{h.c.}\big)\,. \nonumber\\
     &{~}
\end{align}
The connections $\omega_{\mu}{}^{ab}$, $\phi_\mu{}^i$ and $f_{\mu}{}^a$
are algebraically determined by imposing the conventional constraints
\begin{gather}
  R(P)_{\mu \nu}{}^a =  0 ~, \quad
  \gamma^\mu R(Q)_{\mu \nu}{}^i + \tfrac32 \gamma_{\nu}
  \chi^i = 0 ~, \nonumber \\
  e^{\nu}{}_b \,R(M)_{\mu \nu a}{}^b - \mathrm{i} \tilde{R}(A)_{\mu a} +
  \tfrac1{8} T_{abij} T_\mu{}^{bij} -\tfrac{3}{2} D \,e_{\mu a} = 0
  \,.  \label{eq:conv-constraints}
\end{gather}
Their solution is given by
\begin{align}
  \label{eq:dependent}
  \omega_\mu{}^{ab} =&\, -2e^{\nu[a}\partial_{[\mu}e_{\nu]}{}^{b]}
     -e^{\nu[a}e^{b]\sigma}e_{\mu c}\partial_\sigma e_\nu{}^c
     -2e_\mu{}^{[a}e^{b]\nu}b_\nu   
      \, -\ft{1}{4}(2\bar{\psi}_\mu^i\gamma^{[a}\psi_i^{b]}
     +\bar{\psi}^{ai}\gamma_\mu\psi^b_i+{\rm h.c.}) \,,\nonumber\\
     \phi_\mu{}^i  =& \, \tfrac12 \left( \gamma^{\rho \sigma} \gamma_\mu -
    \tfrac{1}{3} \gamma_\mu \gamma^{\rho \sigma} \right) \left(
    \mathcal{D}_\rho
    \psi_\sigma{}^i - \tfrac{1}{16} T^{abij} \gamma_{ab} \gamma_\rho
    \psi_{\sigma j} + \tfrac{1}{4} \gamma_{\rho \sigma} \chi^i \right)
    \,,  \nonumber\\
    f_\mu{}^{\mu}  =& \, \tfrac{1}{6} R(\omega,e) - D - \left(
      \tfrac1{12} e^{-1}
    \varepsilon^{\mu \nu \rho \sigma} \bar{\psi}_\mu{}^i \, \gamma_\nu
    \mathcal{D}_\rho \psi_{\sigma i} - \tfrac1{12} \bar{\psi}_\mu{}^i
    \psi_\nu{}^j T^{\mu \nu}{}_{ij} - \tfrac1{4} \bar{\psi}_\mu{}^i
    \gamma^\mu \chi_i +
    \mbox{h.c.} \right) \, .
\end{align}
We will also need the bosonic part of the expression for the
uncontracted connection $f_\mu{}^a$,
\begin{equation}
  \label{eq:f-bos-uncon}
  f_\mu{}^a= \tfrac12 R(\omega,e)_\mu{}^a -\tfrac14 \big(D+\tfrac13
  R(\omega,e)\big) e_\mu{}^a -\tfrac12\mathrm{i}\tilde R(A)_\mu{}^a +
  \tfrac1{16} T_{\mu b} {}^{ij} T^{ab}{}_{ij} \,,
\end{equation}
where $R(\omega,e)_\mu{}^a= R(\omega)_{\mu\nu}{}^{ab} e_b{}^\nu$ is
the non-symmetric Ricci tensor, and $R(\omega,e)$ the corresponding
Ricci scalar. The curvature $R(\omega)_{\mu\nu}{}^{ab}$ is associated
with the spin connection field $\omega_\mu{}^{ab}$.\\
It is convenient to modify two of the curvatures by
including suitable covariant terms,
\begin{align}
  \label{eq:mod-curv}
  \cR(M)_{ab}{}^{\!cd} =\, &    R(M)_{ab}\,^{cd} + \ft1{16}\big(
  T_{abij}\,T^{cdij} + T_{ab}{}^{ij}\, T^{cd}{}_{ij}  \big)\,, \nonumber \\
  {\cal R}(S)_{ab}{}^i =\, & R(S)_{ab}{}^i + \ft34
  T_{ab}{}^{ij} \chi_j\,,
\end{align}
where we observe that $\gamma^{ab} \big(\mathcal{R}(S)- \mathcal{R}(S)\big)_{ab}{}^i=0$.  The modified curvature
$\cR(M)_{ab}{}^{\!cd}$ satisfies the following relations,
\begin{align}
   \label{eq:RM-constraints}
  \cR(M)_{\mu\nu}{}^{\!ab} \,e^\nu{}_b =\,& \mathrm{i} \tilde
  R(A)_{\mu\nu} e^{\nu a} +\ft32 D\,e_\mu{}^a\,, \nonumber\\
  \ft14 \varepsilon_{ab}{}^{ef} \,\varepsilon^{cd}{}_{gh}\,
 \cR(M)_{ef}{}^{\!gh} =\,& \cR(M)_{ab}{}^{\!cd} \,,\nonumber\\
  \varepsilon_{cdea}\,\cR(M)^{cd\,e}{}_{\!b} =\,&
  \varepsilon_{becd} \,\cR(M)_a{}^{\!e\,cd}  = 2\tilde{R}(D)_{ab} =
  2\mathrm{i} R(A)_{ab}\,.
\end{align}
The first of these relations corresponds to the third constraint given
in~\eqref{eq:conv-constraints}, while the remaining equations follow
from combining the curvature constraints with the Bianchi
identities. Note that the modified curvature does not satisfy the pair
exchange property; instead we have,
\begin{equation}
\cR(M)_{ab}{}^{\!cd} = \cR(M)^{cd}{}_{\!ab} + 4 i
\delta^{[c}_{[a} \, \tilde R(A) _{b]}^{~}{}^{\!d]}\,.
\end{equation}
\vspace{0.4cm}\\
Now that the gauge fields of the superconformal agebra have been introduced, the matter multiplets can be discussed. A covariant version of the Weyl multiplet will also be presented. We refer to the original paper \cite{deWit:1980lyi} for detailed information about the non-linear multiplet.\\

We start from the chiral multiplet which is typically obtained by imposing the lowest component, the scalar $A$, to be a conformal primary (i.e. S-susy invariant) transforming chirally under Q-susy. The superconformal algebra closes on the following chiral multiplet representation:
\begin{equation}
\Phi=(A,\psi_i,B_{ij},F^-_{ab}, \Lambda_i,C) \,,
\end{equation}
 where $A$ and $C$ are complex scalars, $\psi_i$ and $\Lambda_i$ are ${\rm SU}(2)$ doublets of chiral fermions, $B_{ij}$ is a complex ${\rm SU}(2)$ triplet and $F^-_{ab}$ is simply an anti-selfdual Lorentz tensor. Each component of the chiral multiplet is characterized by two numbers, $w$ and $c$, called the Weyl and the chiral weight. For chiral
multiplets these weights are related by $c=-w$. In that case the Weyl
weight of $A$ equals $w$ and the highest-$\theta$ component $C$ has Weyl
weight $w+2$ (the list of all the weights assignments is given in Table \ref{t1}). 
\begin{table}[t]
 \begin{center}
  \begin{tabular}{|c||cccccc|}
   \hline
\hline\noalign{\smallskip} 
  & $A$ & $\psi_i$ & $B_{ij}$ & $F^-_{ab}$  & $\Lambda_i$ & $C$ \\[1mm]
\hline
\noalign{\smallskip} 

$w$ & $w$ & $w+\tfrac12$ & $w+1$ & $w+1$  & $w+\tfrac32$ & $w+2$   \\[1mm]
$c$ & $-w$ & $-w+\tfrac12$ & $-w+1$ & $-w+1$  & $-w+\tfrac32$ & $-w+2$   \\[1mm]
$\gamma_5$  & & $+$ & & & $+$&  \\[1mm] 
\hline
  \end{tabular}
  \vspace{0.3cm}
\caption{Weyl ($w$) and chiral ($c$) weights of the chiral multiplet components. The chirality ($\gamma_5$) of the fermion fields is also indicated.}
\label{t1}
 \end{center}
\end{table} 
 The Q- and S- transformation rules of a generic chiral multiplet read:
    \begin{align}
  \label{eq:conformal-chiral}
  \delta A =&\,\bar\epsilon^i\Psi_i\,, \nonumber\\*[.2ex]
  \delta \Psi_i =&\,2\,\Slash{D} A\epsilon_i + B_{ij}\,\epsilon^j +
  \tfrac12   \gamma^{ab} F_{ab}^- \,\varepsilon_{ij} \epsilon^j + 2\,w
  A\,\eta_i\,,  \nonumber\\*[.2ex]
  \delta B_{ij} =&\,2\,\bar\epsilon_{(i} \Slash{D} \Psi_{j)} -2\,
  \bar\epsilon^k \Lambda_{(i} \,\varepsilon_{j)k} + 2(1-w)\,\bar\eta_{(i}
  \Psi_{j)} \,, \nonumber\\*[.2ex]
  \delta F_{ab}^- =&\,\tfrac12
  \varepsilon^{ij}\,\bar\epsilon_i\Slash{D}\gamma_{ab} \Psi_j+
  \tfrac12 \bar\epsilon^i\gamma_{ab}\Lambda_i
  -\tfrac12(1+w)\,\varepsilon^{ij} \bar\eta_i\gamma_{ab} \Psi_j \,,
  \nonumber\\*[.2ex]
  \delta \Lambda_i =&\,-\tfrac12\gamma^{ab}\Slash{D}F_{ab}^-
   \epsilon_i  -\Slash{D}B_{ij}\varepsilon^{jk} \epsilon_k +
  C\varepsilon_{ij}\,\epsilon^j
  +\tfrac14\big(\Slash{D}A\,\gamma^{ab}T_{abij}
  +w\,A\,\Slash{D}\gamma^{ab} T_{abij}\big)\varepsilon^{jk}\epsilon_k
  \nonumber\\*
  &\, -3\, \gamma_a\varepsilon^{jk}
  \epsilon_k\, \bar \chi_{[i} \gamma^a\Psi_{j]} -(1+w)\,B_{ij}
  \varepsilon^{jk}\,\eta_k + \tfrac12 (1-w)\,\gamma^{ab}\, F_{ab}^-
    \eta_i \,, \nonumber\\*[.2ex]
    \delta C =&\,-2\,\varepsilon^{ij} \bar\epsilon_i\Slash{D}\Lambda_j
  -6\, \bar\epsilon_i\chi_j\;\varepsilon^{ik}
    \varepsilon^{jl} B_{kl}   \nonumber\\*
  &\, -\tfrac14\varepsilon^{ij}\varepsilon^{kl} \big((w-1)
  \,\bar\epsilon_i \gamma^{ab} {\Slash{D}} T_{abjk}
    \Psi_l + \bar\epsilon_i\gamma^{ab}
    T_{abjk} \Slash{D} \Psi_l \big) + 2\,w \varepsilon^{ij}
    \bar\eta_i\Lambda_j \,.
\end{align}
In the above equations, the derivatives $D$ are covariantized with respect to 
the gauge transformations of the superconformal algebra 
appropriate for each field, as shown in \eqref{eq:D-epslon}.\\
\\
The vector multiplet is an example of a reduced chiral multiplet, with components:
\begin{equation}
\cX=(X,\Omega_i,W_\mu,Y_{ij}) \,,
\end{equation}
where $X$ is a complex scalar, $\Omega_i$ is an ${\rm SU}(2)$ doublet of chiral fermions, $Y_{ij}$ is a real ${\rm SU}(2)$ triplet, and $W_\mu$ is a physical gauge field.\\

This multiplet is obtained from the chiral multiplet by imposing a specific constraint on the field representation. Specifically we impose a reality constraint on the  complex ${\rm SU}(2)$ triplet $B_{ij}$, which alone fixes every component of the chiral multiplet in terms of the vector multiplet components. The identifications are as follows:
\begin{align}
\label{constrvect2}
 A\vert_\mathrm{vec}&=X\;,
\nonumber\\
 \psi_i\vert_\mathrm{vec}&=\Omega_i\;,
\nonumber\\
 B_{ij}\vert_\mathrm{vec}&=Y_{ij}=\varepsilon_{ik}\varepsilon_{jl}\,Y^{kl}\;,
\nonumber\\
 F^-_{ab}\vert_\mathrm{vec}&=F^-_{ab}+\tfrac14\,(\bar\psi_\rho{}^i\gamma_{ab}\gamma^\rho\,\Omega^j
+\bar X\, \bar\psi_\rho{}^i\gamma^{\rho\sigma}\gamma_{ab}\psi_\sigma{}^j
+\mathrm{c.c}-\bar X\,T_{ab}{}^{ij})\eps_{ij}\;,
\nonumber\\
 \Lambda_i\vert_\mathrm{vec}&=-\varepsilon_{ij}\,\Slash{D}\Omega^j\;,
\nonumber\\
 C\vert_\mathrm{vec}&=-2\Box_\mathrm{c}\bar X-\tfrac14\hat F^+_{ab}\,T^{ab}{}_{ij}\eps^{ij}-3\bar\chi_i\Omega^i\;,
\end{align}
where the symbol $\Box_\mathrm{c}=D^\mu D_\mu$ is the superconformal d'Alembertian and $F_{ab} $ is used to indicate the abelian field strength $F_{ab}=2\,e_a{}^{[\mu}\,e_b{}^{\nu]}\,\pa_\mu W_\nu$, which satisfies now a (superconformal) Bianchi identity.\\

In Table \ref{t2} we show the Weyl and chiral weight assignment for the vector multiplet components.
\begin{table}[t]
 \begin{center}
  \begin{tabular}{|c||cccc|}
   \hline
\hline\noalign{\smallskip} 
  & $X$ & $\Omega_i$ & $W_\mu$ & $Y_{ij}$ \\[1mm]
\hline
\noalign{\smallskip} 

$w$ & $1$ & $\tfrac32$ & $0$ & $2$   \\[1mm]
$c$ & $-1$ & $-\tfrac12$ & $0$ & $0$  \\[1mm]
$\gamma_5$  & & $+$ & & \\[1mm] 
\hline

  \end{tabular}
    \vspace{0.3cm}
\caption{Weyl ($w$) and chiral ($c$) weights of the vector multiplet components. The chirality ($\gamma_5$) of the fermion field $\Omega_i$ is also indicated.}
\label{t2}
 \end{center}
\end{table}
The Q- and S-supersymmetry transformation rules for the vector multiplet in a conformal background take the form,
\begin{align}
 \label{eq:variations-vect-mult-conf}
 \delta X =&\, \bar{\epsilon}^i\Omega_i \,,\nonumber\\
 \delta\Omega_i =&\, 2 \Slash{D} X\epsilon_i
    +\tfrac12 \varepsilon_{ij} \hat F_{\mu\nu}
  \gamma^{\mu\nu}\epsilon^j +Y_{ij} \epsilon^j+2\,X\,\eta_i
  \,,\nonumber\\
 \delta W_{\mu} = &\,
 \varepsilon^{ij} \bar{\epsilon}_i \big(\gamma_{\mu} \Omega_j
+ 2\,\psi_{\mu j}\,X\big)
 + \varepsilon_{ij} \bar{\epsilon}^i\big( \gamma_{\mu} \Omega^{j}
+ 2\,\psi_\mu{}^j\,\bar X\big) 
\,,\nonumber\\
\delta Y_{ij}  = &\, 2\, \bar{\epsilon}_{(i}
 \Slash{D}\Omega_{j)}
 + 2\, \varepsilon_{ik} \varepsilon_{jl}\,
  \bar{\epsilon}^{(k} \Slash{D}\Omega^{l)} \,.
\end{align}
\\
The last multiplet we want to present is the covariant Weyl multiplet from which the Weyl squared multiplet is obtained. This is another reduced chiral multiplet, and its components read:
\begin{align}
\label{cov-weyl-mult}
A_{ab}\vert_{\bf W}&=T_{ab}{}^{ij}\,\varepsilon_{ij}\,,
\nonumber\\
\psi_{abi}\vert_{\mathbf{W}}&=8\,\varepsilon_{ij} R(Q)_{ab}^j\,,
\nonumber\\
B_{abij}\vert_{\mathbf{W}}&=-8\,\varepsilon_{k(i}R(\cV)_{ab\;\;j)}^{-\;\,k}\,,
\nonumber\\
(F^-_{ab})^{cd}\vert_{\mathbf{W}}&=-8\,\cR(M)_{ab}^{-\,cd}
\nonumber\\
\Lambda_{abi}\vert_{\mathbf{W}}&=8\,(\cR(S)^-_{abi}+\tfrac34\,\gamma_{ab}\Slash{D}\chi_i)
\nonumber\\
C_{ab}\vert_{\mathbf{W}}&=4\,D_{[a}\,D^c T_{b]cij}\,\varepsilon^{ij}-\mathrm{dual}\;.
\end{align}
Note that the Weyl tensor is contained inside the highest independent component $(F^-_{ab})^{cd}$ through the $\cR(M)_{ab}^{-\,cd}$ curvature.\\
By squaring the covariant Weyl multiplet $\mathbf{W}$ a scalar chiral multiplet with $w=2$ is obtained,
\begin{align}
  \label{eq:W-squared}
  A   =&\,(T_{ab}{}^{ij}\varepsilon_{ij})^2\,,\nonumber \\[.2ex]
  \Psi_i  =&\, 16\, \varepsilon_{ij}R(Q)^j_{ab} \,T^{klab}
  \, \varepsilon_{kl} \,,\nonumber\\[.2ex]
  B_{ij}   =&\, -16 \,\varepsilon_{k(i}R({\cal
    V})^k{}_{j)ab} \, T^{lmab}\,\varepsilon_{lm} -64
  \,\varepsilon_{ik}\varepsilon_{jl}\,\bar R(Q)_{ab}{}^k\, R(Q)^{l\,ab}
  \,,\nonumber\\[.2ex]
  F^{-ab}   =&\, -16 \,\cR(M)_{cd}{}^{\!ab} \,
  T^{klcd}\,\varepsilon_{kl}  -16 \,\varepsilon_{ij}\, \bar
  R(Q)^i_{cd}  \gamma^{ab} R(Q)^{cd\,j}  \,,\nonumber\\[.2ex]
 \Lambda_i  =&\, 32\, \varepsilon_{ij} \,\gamma^{ab} R(Q)_{cd}^j\,
  \cR(M)^{cd}{}_{\!ab}
  +16\,({\cal R}(S)_{ab\,i} +3 \gamma_{[a} D_{b]}  \chi_i) \,
  T^{klab}\, \varepsilon_{kl} \nonumber\\
  &\, -64\, R({\cal V})_{ab}{}^{\!k}{}_i \,\varepsilon_{kl}\,R(Q)^{ab\,l}
  \,,\nonumber\\[.2ex]
  C  =&\,  64\, \cR(M)^{-cd}{}_{\!ab}\,
 \cR(M)^-_{cd}{}^{\!ab}  + 32\, R({\cal V})^{-ab\,k}{}_l^{~} \,
  R({\cal V})^-_{ab}{}^{\!l}{}_k  \nonumber \\
  &\, - 32\, T^{ab\,ij} \, D_a \,D^cT_{cb\,ij} +
  128\,\bar{\mathcal{R}}(S)^{ab}{}_i  \,R(Q)_{ab}{}^i  +384 \,\bar
  R(Q)^{ab\,i} \gamma_aD_b\chi_i   \,.
\end{align}
Both the covariant Weyl multiplet $\mathbf{W}$ and its square are functions of the curvatures of the local superconformal algebra. As expected from a reduced chiral multiplet, the highest components of the Weyl multiplet are not independent.

\bibliography{references}{}
\bibliographystyle{myieeetreprint}

\end{document}